\documentclass[10pt,twocolumn]{IEEEtran}
\usepackage[T1]{fontenc}
\ifCLASSINFOpdf
\else
\fi
\usepackage{amsmath}
\usepackage{amsbsy}
\usepackage{amssymb}
\usepackage{amsthm}
\usepackage{cite}
\usepackage{graphics}
\usepackage{epsfig}
\usepackage{graphicx}
\usepackage{psfrag}
\usepackage{color}
\usepackage{cases}
\usepackage{bm}
\newtheorem{theorem}{Theorem}

\newtheorem{corollary}{Corollary}
\DeclareMathOperator*{\argmax}{arg\,max}
\newlength{\figwidth}
\ifCLASSOPTIONcompsoc
\usepackage[caption=false,font=normalsize,labelfont=sf,textfont=sf]{subfig}
\else
\usepackage[subrefformat=parens,labelformat=parens,caption=false,font=footnotesize]{subfig}
\fi

\usepackage{url}
\IEEEoverridecommandlockouts
\begin{document}
\title{Simple Approximations of the SIR Meta Distribution in General Cellular Networks}
\author{Sanket~S.~Kalamkar,~\IEEEmembership{Member,~IEEE,}
       and~Martin~Haenggi,~\IEEEmembership{Fellow,~IEEE}
\thanks{S. S. Kalamkar is with INRIA, Paris, France. M. Haenggi is with the Department of Electrical Engineering, University of Notre Dame, Notre Dame, IN, 46556 USA. (e-mail: $\lbrace$skalamka, mhaenggi$\rbrace$@nd.edu).}
\thanks{This work was done when S. S. Kalamkar was with the University of Notre Dame, IN, USA.}
\thanks{This work is supported by the US National Science Foundation (grant CCF 1525904).}
\thanks{Part of this work was presented at the 2018 IEEE International Conference on Communications (ICC'18)~\cite{sanket_icc18}.}}
\maketitle

\begin{abstract}

Compared to the standard success (coverage) probability, the meta distribution of the signal-to-interference ratio (SIR) provides much more fine-grained information about the network performance. We consider general heterogeneous cellular networks (HCNs) with base station tiers modeled by arbitrary stationary and ergodic non-Poisson point processes. The exact analysis of non-Poisson network models is notoriously difficult, even in terms of the standard success probability, let alone the meta distribution. Hence we propose a simple approach to approximate the SIR meta distribution for non-Poisson networks based on the ASAPPP (``approximate SIR analysis based on the Poisson point process'') method. We prove that the asymptotic horizontal gap $G_0$ between its standard success probability and that for the Poisson point process exactly characterizes the gap between the $b$th moment of the conditional success probability, as the SIR threshold goes to $0$. The gap $G_0$ allows two simple approximations of the meta distribution for general HCNs: 1) the {\em per-tier} approximation by applying the shift $G_0$ to each tier and 2) the {\em effective gain} approximation by directly shifting the meta distribution for the homogeneous independent Poisson network. Given the generality of the model considered and the fine-grained nature of the meta distribution, these approximations work surprisingly well.

\end{abstract}
\IEEEpeerreviewmaketitle

\section{Introduction}
\subsection{Motivation and Objective}
The accurate modeling of base station (BS) locations is important to characterize the performance of cellular networks and obtain useful design insights. Traditionally, in a cellular network, the BS locations were modeled in a deterministic (regular) manner using either triangular or square lattices. The lattice model has been extensively studied using simulations since it is usually analytically intractable. However, to meet an exponential growth in mobile traffic and improve the spatial reuse, the deployment of cellular networks has become more irregular and heterogeneous. For example, in a geographical region, macro, pico, and femto BSs can coexist. The network tiers possess different characteristics such as different BS densities, different path loss exponents, and different deployment structures ({\em e.g.}, clustered or repulsive deployments). 

The Poisson point process (PPP) may be used to model irregular and real-world BS deployments~\cite{anjin_cell_model}. The modeling of BS locations by the PPP has become popular due to its analytical tractability, which leads to crisp insights about the network performance. However, in an actual cellular network, the BS deployment is neither completely random (as the PPP) nor completely regular (as the triangular and square lattices)---it lies somewhere in between. The BS deployment depends heavily on the topology and the type of the geographical region (urban or rural). As a result, a single point process model may not be applicable in all scenarios. For example, using actual data from the UK, it is shown in~\cite{anjin_cell_model} that there exists repulsion among BSs, which can be modeled using hard-core point processes~\cite[Chapter 3]{martin_book}. On the other hand, in~\cite{lee_cell_model}, the Poisson cluster process~\cite[Chapter 3]{martin_book} is shown to accurately model the BS deployment in many cities. Especially, at a larger geographical scale, the BSs appear to form a cluster point process due to the high density in urban regions and low density in rural regions. Hence it is important to investigate the performance of non-Poisson cellular networks. 

The main impediment to the study of non-Poisson cellular networks is that, compared to the PPP, their analysis is much harder due to the dependence between the BS locations. Thus it would be convenient if the performance of non-Poisson cellular networks could be related (approximately) to that of Poisson cellular networks.

Recently, in~\cite{martin_meta_2016}, a new fundamental performance metric called the \textit{meta distribution} of the signal-to-interference ratio (SIR) is introduced for cellular networks. The meta distribution, defined as the distribution of the conditional success probability given the point process, is an important performance metric as it answers a key question: ``How are the individual link success (or coverage) probabilities\footnote{The success probability of a link is the probability that the SIR at the receiver of that link is greater than the target SIR threshold $\theta$.} distributed in a realization of the cellular network?'' The answer directly leads to the performance of the ``5\% user,'' which corresponds to the performance of the top $95\%$ of users and is an important design criterion for cellular operators. The meta distribution provides much more fine-grained information about the network than the standard success (coverage) probability; the latter provides just the average of individual link success probabilities in each realization of the network and thus yields limited information about the network. In contrast, the meta distribution provides the distribution of the link success probability conditioned on the point process and thus allows the analysis of the network at a finer level.

The goal of this paper is to study the meta distribution in heterogeneous non-Poisson cellular networks, where the cellular networks consist of multiple tiers and the BSs in each tier may form an arbitrary stationary and ergodic point process. To achieve this goal, we have to overcome two main difficulties:
\begin{itemize}
\item[1)] The direct calculation of the meta distribution seems infeasible even for the Poisson cellular network~\cite{martin_meta_2016}---one has to calculate the moments of the conditional success probability and then use the Gil-Pelaez theorem~\cite{gp_theorem} to calculate the meta distribution.
\item[2)] The analysis for non-Poisson cellular networks is significantly more difficult than that for Poisson cellular networks. In fact, obtaining an analytical expression of the (standard) success probability is extremely difficult in non-Poisson networks. Even for the arguably second-simplest model, the Ginibre point process, it can only be given in the form of an expression in which 3 integrals, an infinite product, and an infinite sum are nested~\cite[Theorem 2]{deng_gpp}.
\end{itemize} 
Consequently, analyzing the meta distribution in non-Poisson networks is very challenging. This problem becomes even worse in the case of heterogeneous cellular networks, where one needs to consider the intra-tier as well as the inter-tier interference. In this paper, we propose two simple and indirect approaches to approximately calculate the meta distribution for general heterogeneous cellular networks (HCNs) by comparing it to that for a homogeneous independent PPP (HIP) model where the BSs in each tier are modeled by an independent and homogeneous PPP. 

\subsection{Related Work}
The works in \cite{Dhillon_HCN,Mukherjee_HCN,Nigam_HCN} obtained analytically tractable results for the HIP model. For HCNs with non-Poisson deployments, it is often the case that it is hard to perform an exact mathematical analysis of key performance metrics such as the SIR distribution (sometimes called the coverage probability). Even if an exact expression of the SIR distribution exists, it is available in a complex form that does not help gain insights about the performance of the network for different network parameters~\cite{NAKATA20147,Deng_HCN_dependence,Suryaprakash_HCN,Flint_HCN,saha_3gpp}. 

Fortunately, \cite{anjin_cell_model} observed that the SIR distribution for the downlink of cellular networks modeled by different non-Poisson point processes can be closely approximated by simply applying a horizontal shift to the SIR distribution curve for the PPP model. The approximation becomes asymptotically exact as the SIR threshold $\theta \to 0$~\cite{anjin_deployment_gain}. The horizontal shift is termed the deployment gain in~\cite{anjin_deployment_gain} since the shift is because of the deployment. This method of approximating the SIR distribution for a non-Poisson point process model by that for the PPP model is called ``Approximate SIR analysis based on the PPP'' (ASAPPP) in~\cite{martin_asappp}. \cite{martin_misr} showed that the deployment gain as $\theta \to 0$ can be expressed as the ratio of the mean interference-to-signal ratio (MISR) of the two different point processes under consideration. Further, \cite{ganti_asymptotics} proved that the deployment gain as $\theta \to \infty$ is determined by the expected fading-to-interference ratio (EFIR). A key observation from~\cite{ganti_asymptotics} is that the deployment gain as $\theta \to 0$, denoted by $G_0$, provides an excellent approximation to the entire SIR distribution. In~\cite{takayashi_gain}, a formula is derived to approximately calculate $G_0$ analytically for (stationary) non-Poisson point process models whenever the second moment of the contact distance is available. \cite{mh_general_asappp} showed that the ASAPPP approximation works very well for HCNs with non-Poisson deployment of BSs. First, \cite{mh_general_asappp} proposed the per-tier ASAPPP-based approximation for general HCNs, where the ASAPPP approximation was used to approximate the SIR distribution corresponding to each non-Poisson tier using the MISR-based deployment gain $G_0$. Second, when the path loss of each tier was the same, \cite{mh_general_asappp} showed that the SIR distribution for general HCNs could be directly obtained from that for the HIP model by scaling the SIR threshold by the effective gain. 

The meta distribution of the SIR for cellular networks was proposed in~\cite{martin_meta_2016}, where the focus was on the downlink of the Poisson cellular network. Furthermore, the meta distribution of the SIR was calculated for both the downlink and the uplink of the Poisson cellular network with power control in~\cite{yuanjie}, for the downlink  Poisson cellular network underlaid with a device-to-device (D2D) network in~\cite{martin_d2d}, for the non-orthogonal multiple access (NOMA) network in~\cite{MD_NOMA}, and with base station cooperation in~\cite{cui_tcom}. For general cellular networks with a multi-slope path loss model, \cite{afshang_md} gave a scaling law involving the parameters of BS and user point processes ({\em e.g.}, the density of the point process) that keeps the meta distribution of the SIR the same. For the HIP-based $K$-tier HCN, \cite{Yuanjie_HCN} calculated the SIR meta distribution with cell range expansion. 

Overall, for general HCNs with non-Poisson deployment of BSs, the focus has been only on the SIR distribution, and the SIR meta distribution is calculated only for Poisson cellular networks. In this paper, we determine the fine-grained network performance of general HCNs through the SIR meta distribution. Since the model for cellular networks considered is quite general, we cannot expect to obtain exact analytical expressions for the SIR meta distribution. Hence we propose simple approximations that enable a quick calculation of the SIR meta distribution for general HCNs. Note that, until now, the ideas of the SIR meta distribution and the ASAPPP have been explored and applied separately. Applying the shift to the SIR distribution (which is just the mean of the SIR meta distribution) does not imply that the shifting approach also works for the entire SIR meta distribution. That said, combining these ideas is of significant importance because ``mean to distribution'' and ``Poisson to non-Poisson'' are highly non-trivial extensions.

\begin{table*}
\caption{Notation and Abbreviation} \label{tab:notation}
\begin{center}
\renewcommand{\arraystretch}{1.05}
\begin{tabular}{|c | p{11cm}| }
\hline 
 {Notation} & {\hspace{2cm}}{Definition/Meaning}
\\
\hline
\hline 
$\Phi_k$ & Point process of BSs of $k$th tier \\
$\Phi_k^{\rm PPP}$ & Approximation of the point process of BSs of $k$th tier by a PPP of the same density as $\Phi_k$\\
$\lambda_k$ & Density of BSs of $k$th tier  \\
$P_k$ & Transmit power of BSs of $k$th tier\\
$[K]$ & $\lbrace 1, 2, \dotsc, K \rbrace$\\
$\mathsf{SIR}$ & Signal-to-interference ratio \\
$\theta$ & SIR threshold\\
$\alpha_k$ & Path loss exponent of $k$th tier\\
$\delta_k$ & $2/\alpha_k$\\
$G_0$ & Asymptotic SIR gain as $\theta \to 0$\\
$G_k$ & $G_0$ for $k$th tier, $k \in [K]$\\ 
$G_{\rm eff}$ & Effective gain for HCNs\\
$p_{\rm s}(\theta)$ & Standard success (coverage) probability\\
$P_{\rm s}(\theta)$ & Conditional link success probability\\
$M_{b}(\theta)$ & $b$th moment of the conditional link success probability\\
$\bar{F}(\theta, x)$ & SIR meta distribution for the target reliability of $x$ \\
$\mathsf{MISR}$ & Mean interference-to-signal ratio\\
$\mathsf{EFIR}$ & Expected fading-to-interference ratio\\
$R_{\rm pert}$ & Perturbation radius for the perturbed triangular lattice \\
$p, u$ & Probability of one-point cluster and the distance between two points in a two-point cluster for the Gauss-Poisson point process\\
$\lambda_{\rm p}$, $\bar{c}$, $r_{\rm c}$ & Density of parent point process, the mean number of points in a cluster, and the radius of the cluster for the Mat{\'e}rn cluster process\\
\hline 
\end{tabular}
\end{center}
\end{table*}

\subsection{Contributions}
This paper makes the following contributions:
\begin{itemize}
\item[1)] For cellular networks, we apply the idea of ASAPPP to the meta distribution and propose a simple and novel method, called AMAPPP which stands for ``Approximate meta distribution analysis using the PPP,'' to obtain the meta distribution for an arbitrary stationary and ergodic point process from the meta distribution for the PPP.
\item[2)] We prove that, as $\theta \to 0$, the $b$th moment of the conditional success probability for a stationary and ergodic point process model can be obtained exactly by shifting that for the PPP model by the asymptotic deployment gain $G_0$.
\item[3)] For Rayleigh fading and an unbounded path loss model, we confirm by simulations that applying the horizontal shift by the gain $G_0$ to the meta distribution for the PPP closely approximates the meta distribution for the stationary triangular lattice, the perturbed triangular lattice, the Gauss-Poisson point process, and the Mat{\'e}rn cluster process (two regular and two cluster point processes).
\item[4)] We extend the AMAPPP approach to general HCNs, which approximates the $b$th moment of the conditional success probability corresponding to each tier using the MISR-based gain $G_0$. This {\em per-tier} approach is further used to approximately calculate the meta distribution for general HCNs.
\item[5)] When the path loss exponents in all tiers are the same, we obtain an {\em effective gain} using the MISR-based gain $G_0$ corresponding to each tier, which can be simply applied to shift the meta distribution for the HIP cellular network to approximately obtain the meta distribution for general HCNs.
\end{itemize}

\subsection{Organization of the Paper}
The rest of the paper is organized as follows. In Sec.~\ref{sec:system_model}, we provide the network model, describe the point process models considered in this paper, and briefly review the SIR meta distribution and the ASAPPP method. In Sec.~\ref{sec:AMAPPP_single}, we propose the AMAPPP approach for single-tier non-Poisson cellular networks, which is extended to general HCNs in Sec.~\ref{sec:AMAPPP_HCN}. We provide conclusions in Sec.~\ref{sec:conclusions}.

\section{System Model}
\label{sec:system_model}
\subsection{Network Model}
We consider a general $K$-tier HCN where the locations of BSs of the $k$th tier are modeled by arbitrary stationary, ergodic, and independent point processes $\Phi_k \subset \mathbb{R}^2$, $k = 1, 2, \dotsc, K$. The density of $\Phi_k$ is $\lambda_k$. All BSs are always active. A base station belonging to $\Phi_k$ transmits at power $P_k$. Due to the stationarity of the point processes, we focus on the cellular user situated at the origin $o = (0, 0)$, henceforth called the typical user. We focus on the downlink with the (on average) strongest-BS association, where the typical user connects to the BS with the strongest received power on average. The other BS transmissions from the same tier as that of the serving BS and those from different tiers cause interference at the typical user. The signal propagation experiences fading as well as path loss. We assume independent and identically distributed (i.i.d.) Rayleigh fading where the channel power gains are exponentially distributed with mean $1$. The path loss function corresponding to the tier $\Phi_k$ is given by $\ell(x) = \|x\|^{-\alpha_k}$, where $\alpha_k > 2$ is the path loss exponent.

We focus on an interference-limited network where the received SIR determines the network performance. Let $x_0 \triangleq \argmax\{x \in \Phi_k, k\in[K] \colon P_k \|x\|^{-\alpha_k}\}$ be the serving BS of the typical user, where $[K] \triangleq \lbrace 1, 2, \dotsc, K \rbrace$. Also let $\Phi_k^{!}$  and $[K]^{!}$ denote $\Phi_k \setminus \lbrace x_0 \rbrace$ and $[K]\setminus\lbrace k \rbrace$, respectively. When the typical user connects to a BS of $\Phi_k$, the SIR at the typical user is given by
\begin{align}
\mathsf{SIR} &\triangleq \frac{S}{I_{\rm Intra} + I_{\rm Inter}} \nonumber \\
&=  \frac{P_k h_{x_0} \|x_0\|^{-\alpha_k}}{\displaystyle\sum_{x \in \Phi_k^{!}}P_k h_x \|x\|^{-\alpha_k} + \sum_{i \in [K]^{!}}\sum_{y \in \Phi_i}P_i h_y \|y\|^{-\alpha_i}},
\label{eq:sir}
\end{align}
where $h_x$ represents the exponential random variable corresponding to the channel power gain between the BS at $x$ and the typical user, $S \triangleq P_k h_{x_0} \|x_0\|^{-\alpha_k}$ is the received signal power at the typical user, $I_{\rm Intra} \triangleq \sum_{x \in \Phi_k^{!}}P_k h_x \|x\|^{-\alpha_k}$ and $I_{\rm Inter} \triangleq \sum_{i \in [K]^{!}}\sum_{y \in \Phi_i}P_i h_y \|y\|^{-\alpha_i}$ denote the interference power received by the typical user from the intra- and inter-tier BS transmissions, respectively. 

The standard success probability is given as
\begin{align}
p_{\rm s}(\theta) &\triangleq \mathbb{P}(\mathsf{SIR} > \theta) \nonumber \\
&= \sum_{k \in [K]} \mathbb{P}(\mathsf{SIR} > \theta, x_0 \in \Phi_k),
\label{eq:suc_prob_het}
\end{align}
where $\theta$ is the SIR threshold.

\subsection{Point Process Models}
In this subsection, we describe four stationary and ergodic point processes, two regular (repulsive) and two cluster point processes, that may be used to model BS locations. These four point processes are used as illustrative examples of stationary and ergodic point processes.
\subsubsection{Stationary Triangular Lattice (TL)}
The triangular lattice is the most regular point process. The stationary triangular lattice is obtained by randomly translating the non-stationary triangular lattice $\mathbb{L} \subset \mathbb{R}^2$ given by
\begin{align}
\mathbb{L} \triangleq \lbrace v \in \mathbb{Z}^2 \colon \boldsymbol Gv \rbrace,
\end{align}
where $\boldsymbol{G} = \eta
\begin{bmatrix}
    1       & 1/2\\
    0       & \sqrt{3}/2
\end{bmatrix}$ 
is the generator matrix and $\eta$ is the distance between any two neighboring points of $\mathbb{L}$. The density of the triangular lattice is $2/(\sqrt{3}\eta^2)$. We translate $\mathbb{L}$ by a random vector $X$ distributed uniformly over the Voronoi cell of the origin to obtain the stationary triangular lattice $\Phi$ as\footnote{An (easier) alternative to generate the stationary triangular lattice is to generate the stationary square lattice, which has square Voronoi cells, and then multiply its points by the generator matrix $\boldsymbol{G}$.}
\begin{align}
\Phi \triangleq \lbrace v \in \mathbb{Z}^2 \colon \boldsymbol{G}v + X \rbrace.
\end{align}
In the rest of the paper, by the ``triangular lattice,'' we mean the ``stationary triangular lattice.''

\subsubsection{Perturbed Triangular Lattice (PTL)}
The second regular point process that we consider is the perturbed triangular lattice (PTL)~\cite[Chapter 2]{martin_book}. The PTL is obtained by perturbing the stationary triangular lattice, {\em i.e.},
\begin{align}
\Phi \triangleq \lbrace v \in \mathbb{Z}^2 \colon \boldsymbol{G}v + X + Y_v\rbrace,
\end{align}
where $(Y_{v})_{v \in \mathbb{Z}^2}$ is a family of i.i.d. random variables. In this paper, we assume that $(Y_v)$ are uniformly distributed on the disk $b(o, R_{\rm pert})$ centered at the origin with radius $R_{\rm pert}$.

\subsubsection{Gauss-Poisson Point Process (GaPPP)}

The GaPPP is a Poisson cluster process where each cluster contains either one or two points with probabilities $p$ and $1-p$, respectively~\cite{Anjin_GaPPP}. For a one-point cluster, the point is at the parent point location, while for a two-point cluster, one of the points is at the parent point location and the other is located at a deterministic distance $u$ from the parent point in a random direction, {\em i.e.}, the second point lies uniformly at random on the circle of radius $u$ centered at the parent point location. 

For a Poisson parent point process $\Phi_{\rm p}$ with density $\lambda_{\rm p}$, let $\Phi_x$ with $x \in \Phi_{\rm p}$ be the clusters of the GaPPP, which are denoted by
\begin{equation}
  \Phi_{x} =
    \begin{cases}
      \lbrace x \rbrace & \text{with probability $p$}\\
      \lbrace x, x+u_x \rbrace & \text{with probability $1-p$},
    \end{cases}      
\end{equation}
where $u_x = (u\sin \phi_x, u\cos \phi_x)$ with $\phi_x$ uniformly distributed in $[0, 2\pi]$. The density of the GaPPP is $\lambda_{\rm p}(2-p)$.

\subsubsection{Mat{\'e}rn Cluster Process (MCP)}

The MCP is a doubly Poisson cluster process, where the parent point process $\Phi_{\rm{p}}$ is a PPP with density $\lambda_{\rm{p}}$ and the daughter points are uniformly distributed within a ball of radius $r_{\rm{c}}$ with each parent point $x_{\rm{p}} \in \Phi_{\rm{p}}$ as its center~\cite[Chapter 3]{martin_book}. The density of the daughter point process of parent $x_{\rm{p}}$ is given by
\begin{align}
\lambda_{\rm{d}}(x) = \frac{\bar{c}}{\pi r_{\rm{c}}^2}\boldsymbol{1}_{B(x_{\rm{p}}, r_{\rm{c}})}(x),
\end{align}
where $B_{(x_{\rm{p}}, r_{\rm{c}})}(x) \triangleq \{x \in \mathbb{R}^2 \colon \|x - x_{\rm{p}}\| \leq r_{\rm{c}}\}$, $\bar{c}$ is the average number of daughter points in a cluster, and $\boldsymbol{1}(\cdot)$ is the indicator function. The density of the MCP is $\lambda = \lambda_{\rm{p}}\bar{c}$.

\subsection{The SIR Meta Distribution}
For an SIR threshold $\theta$ and a reliability threshold $x$, the meta distribution of the SIR is given by 
\begin{align}
\bar{F}(\theta, x) = \bar{F}_{P_{\rm{s}}}(\theta, x) \triangleq \mathbb{P}(P_{\rm{s}}(\theta)> x), \quad \theta \in \mathbb{R}^{+}, x \in [0, 1],
\label{eq:meta_def}
\end{align}
where $P_{\rm{s}}(\theta)$ is a random variable that represents the link success probability conditioned on the point process $\Phi$, given by
\begin{align}
P_{\rm{s}}(\theta) \triangleq \mathbb{P}(\mathsf{SIR} > \theta \mid \Phi).
\end{align}
Here the probability is taken with respect to the fading. The SIR is calculated at the receiver of the link under consideration. The meta distribution is the complementary cumulative distribution function (ccdf) of the conditional link success probability $P_{\rm{s}}(\theta)$. Interpreted differently, for a stationary and ergodic point process, the SIR meta distribution yields the fraction of cellular users that achieve an SIR of $\theta$ with reliability at least $x$. Also, note that the meta distribution of the rate $R$ can be obtained from the meta distribution of the SIR using the relation $R = W \log_2(1 + \mathsf{SIR})$ with $W$ denoting the bandwidth~\cite{deng_energy}.

The standard success (coverage) probability $p_{\rm{s}}(\theta)$ (the SIR distribution) can be obtained from the SIR meta distribution as the mean of the conditional link success probability $P_{\rm{s}}(\theta)$, \textit{i.e.},
\begin{align}
p_{\rm{s}}(\theta) \triangleq \mathbb{P}(\mathsf{SIR} > \theta) = \mathbb{E}(P_{\rm{s}}(\theta)) = \int_{0}^{1} \bar{F}(\theta, x)\mathrm{d}x.
\label{eq:suc_prob}
\end{align}
Clearly, the distribution of $P_{\rm{s}}(\theta)$ provides much more fine-grained information than merely its average $p_{\rm{s}}(\theta)$. 
\subsubsection{Exact Calculation of the Meta Distribution}
Finding the exact meta distribution directly seems infeasible, but if we can calculate the moments of the conditional link success probability $P_{\rm s}(\theta)$
\begin{align}
M_b(\theta) \triangleq \mathbb{E}(P_{\rm{s}}(\theta)^b), \quad b\in \mathbb{C},
\end{align}
as has been done in~\cite{martin_meta_2016} for Poisson cellular networks, we can calculate the exact meta distribution in \eqref{eq:meta_def} using the Gil-Pelaez theorem~\cite{gp_theorem} as
\begin{align}
\bar{F}(\theta, x)&=  \frac{1}{2}+\frac{1}{\pi} \int\limits_0^\infty \frac{\Im(e^{-jt \log x}M_{jt})}{t}\mathrm{d}t ,
\label{eq:exact_meta}
\end{align}
where $\Im(u)$ is the imaginary part of $u \in \mathbb{C}$. Even though the expression of the meta distribution in \eqref{eq:exact_meta} is exact, its complexity makes it hard to gain direct insights, and it is not very convenient to evaluate numerically.\footnote{To numerically evaluate the integral in \eqref{eq:exact_meta}, one needs to carefully select the appropriate numerical integration range and the step size~\cite{martin_meta_calculation}.}

With the nearest-BS association and the path loss exponent $\alpha$, for the downlink of a single-tier Poisson cellular network and $K$-tier HIP model, the $b$th moment $M_b^{\rm{PPP}}(\theta)$ of the conditional link success probability is simply given by~\cite{martin_meta_2016}
\begin{align}
M_b^{\rm{PPP}}(\theta) = \frac{1}{\:_2F_1(b, -\delta; 1-\delta; -\theta)}, \quad b \in \mathbb{C},
\label{eq:Mb_PPP}
\end{align}
where $_2F_1(\cdot,\cdot;\cdot;\cdot)$ denotes the Gauss hypergeometric function and $\delta \triangleq 2/\alpha$. On the other hand, in a non-Poisson cellular network, the calculation of $M_b(\theta)$---let alone the meta distribution---is quite difficult. Hence it would be extremely useful to have a simple approximation to obtain $M_b(\theta)$ and the meta distribution in a non-Poisson cellular network. We propose to do so by combining the meta distribution with the ASAPPP approach, discussed in the next subsection.

\subsubsection{Approximate Calculation of the Meta Distribution}
For different network settings, the meta distribution can be accurately approximated by the beta distribution by matching the first and the second moments of the conditional link success probability~\cite{martin_meta_2016,yuanjie,martin_d2d,cui_tcom,Yuanjie_HCN}.

\subsection{Approximate SIR Analysis based on the PPP (ASAPPP)}
\subsubsection{Single-tier Network}
ASAPPP is the method that provides an approximation of the SIR distributions in non-Poisson networks by that in the Poisson network by applying a horizontal shift to the latter. This method asserts that if the network model under consideration and the Poisson model only differ in the type of the underlying point process, then the SIR ccdf for the network model under consideration can be closely approximated using the SIR ccdf for the PPP by scaling the SIR threshold $\theta$ by a certain factor $G_0$~\cite{ganti_asymptotics}, \textit{i.e.},
\begin{align}
p_{\rm{s}}(\theta) \approx p_{\rm{s}}^{\rm{PPP}}(\theta/G_0),
\end{align}
which corresponds to a horizontal shift by $G_0$ (in dB) if $\theta$ is plotted in dB. The subscript in $G_0$ corresponds to $\theta \to 0$, \textit{i.e.}, the shift is calculated for $\theta \to 0$. This asymptotic shift $G_0$ can also be interpreted as an SIR gain, similar to the notion of the coding gain in coding theory~\cite{costello}. As shown in~\cite{ganti_asymptotics}, the gain $G_0$ provides an excellent approximation to the entire SIR distribution. This approximation becomes exact as $\theta \to 0$, \textit{i.e.},
\begin{align}
p_{\rm{s}}(\theta) \sim p_{\rm{s}}^{\rm{PPP}}(\theta/G_0), \quad \theta \to 0.
\end{align}
Moreover, the gain $G_0$ shows little sensitivity to the path loss exponent or the fading model~\cite{ganti_asymptotics}; it is a robust constant that captures the difference in the network topologies due to the underlying point process models. The gain $G_0$ can be expressed using the mean interference-to-signal ratios (MISRs) of the point process $\Phi$ under consideration and the PPP as~\cite{martin_misr}
\begin{align}
G_0 &= \frac{\mathsf{MISR}_{\rm{PPP}}}{\mathsf{MISR}},
\label{eq:G_0} \\
& = \frac{2}{\alpha -2}\frac{1}{\mathsf{MISR}},
\label{eq:G_0_ppp}
\end{align}
where $\alpha$ is the path loss exponent for a single-tier network, and
\begin{align}
\mathsf{MISR} &\triangleq \mathbb{E}\left(\frac{\sum_{x \in \Phi \setminus \{x_0\}} h_x \|x\|^{-\alpha}}{\|x_0\|^{-\alpha}}\right) \nonumber \\
&= \mathbb{E}\left(\frac{\sum_{x \in \Phi \setminus \{x_0\}} \|x\|^{-\alpha}}{\|x_0\|^{-\alpha}}\right)
\label{eq:MISR}
\end{align}
and $\mathsf{MISR}_{\rm{PPP}} = 2/(\alpha - 2)$ are the MISRs of the network model under consideration and the PPP, respectively.  The numerical calculation of $G_0$ is quite easy since it just depends on the network geometry.\footnote{For a stationary non-Poisson point process, an analytical expression of the MISR (and hence that of the gain $G_0$) is currently unavailable. An approximate formula for $G_0$ is obtained in a very recent paper~\cite{takayashi_gain}.} This motivates us to investigate whether a horizontal shift of $G_0$ to the meta distribution for the Poisson cellular network approximates the meta distribution for a non-Poisson cellular network. We show that this is indeed the case. We call this approach ``Approximate meta distribution analysis using the PPP'' (AMAPPP). 

\subsubsection{K-tier General Heterogeneous Network}

In a general HCN, as given by~\eqref{eq:suc_prob_het}, the overall standard success probability is the sum of the probabilities of the joint events that $\mathsf{SIR} > \theta$ and that the typical user is served by the $k$th tier where $k \in [K]$.

When the typical user connects to a tier modeled by a stationary non-Poisson point process, the tier is treated as a PPP while shifting the SIR threshold $\theta$ to $\theta/G_0$ in the SIR distribution. Also, the interference from other tiers modeled by stationary non-Poisson point processes is approximated by that from a PPP. Such an approximation is called the ``per-tier ASAPPP'' in \cite{mh_general_asappp} since each tier is treated ``as a PPP'' and is shown to be quite accurate for the entire SIR distribution. We extend the per-tier ASAPPP approach to the meta distribution, which we call the ``per-tier AMAPPP'' approach. It is discussed in Sec.~\ref{sec:per_tier_AMAPPP}.

\section{AMAPPP Approach for Single-tier Networks}
\label{sec:AMAPPP_single}
In this section, we focus on single-tier cellular networks, where the BS deployment follows a stationary and ergodic non-Poisson point process. 

\begin{figure*}
  \centering
  \subfloat[TL, $G_{0}= 3.6099$ dB.\label{fig:Mb_TL_PPP_single}]{\includegraphics[scale=0.44]{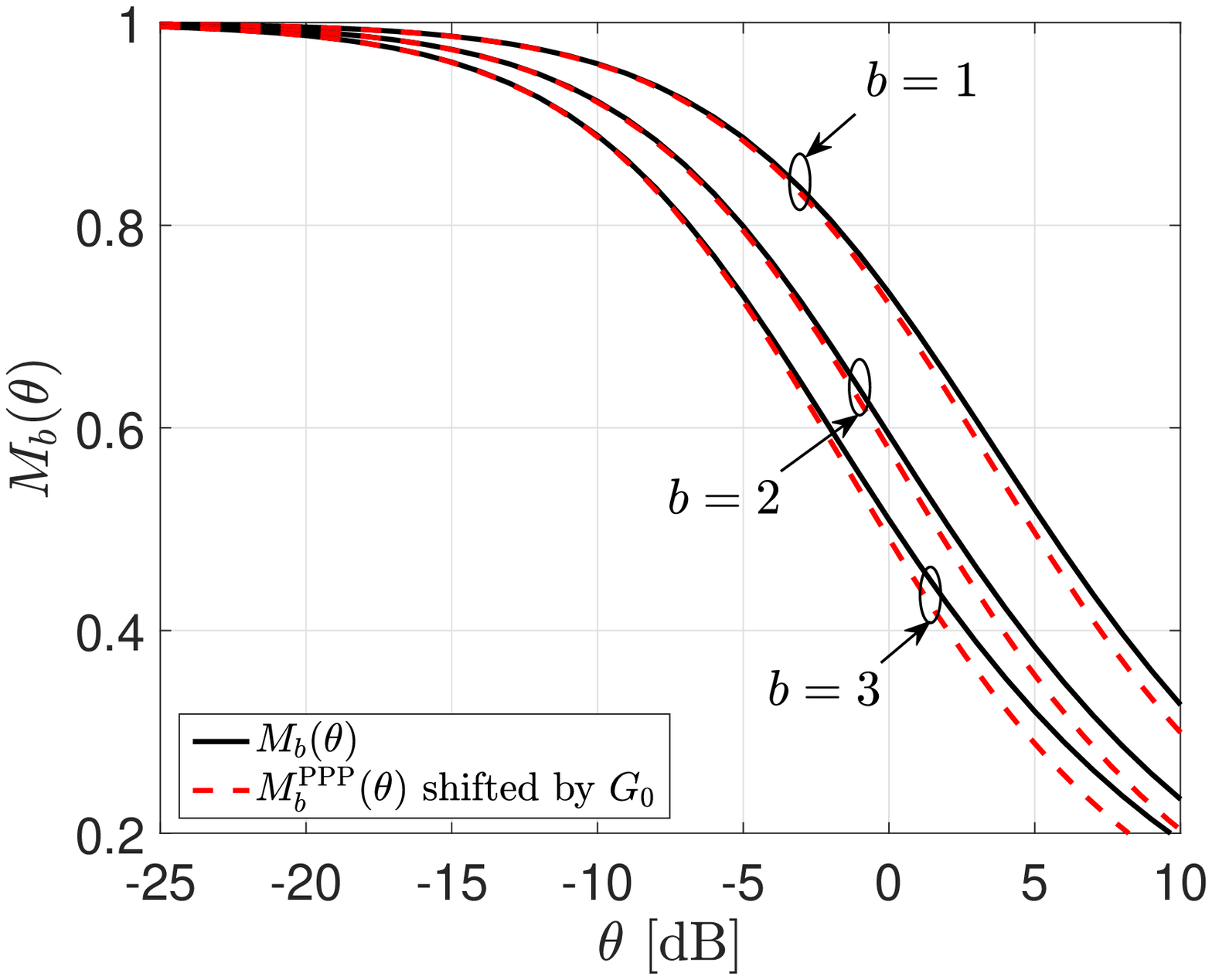}}
  \subfloat[PTL, $G_{0}= 1.8343$ dB\label{fig:Mb_PTL_PPP_single}]{\includegraphics[scale=0.44]{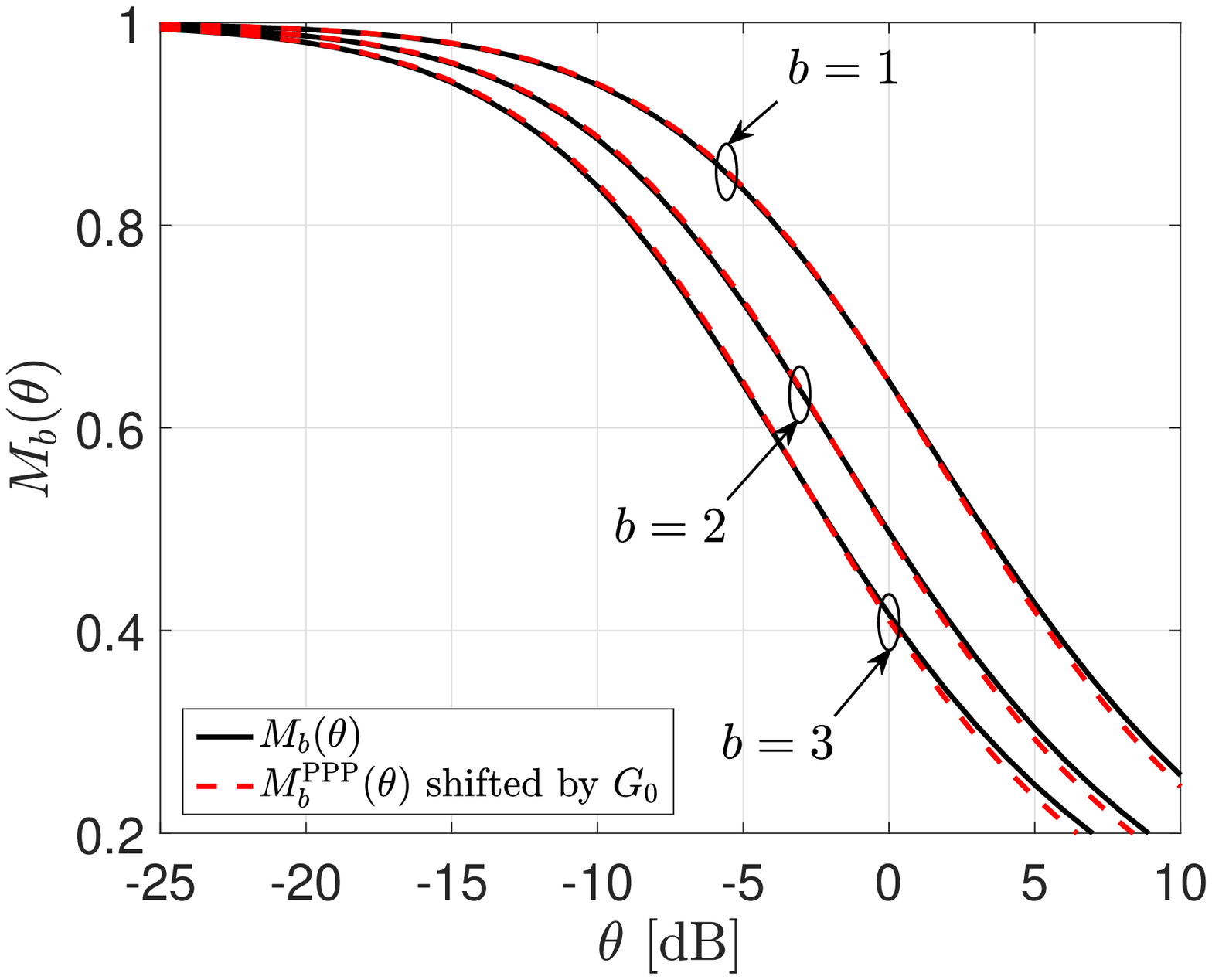}}\\
   \subfloat[GaPPP, $G_{0}= -1.3768$ dB\label{fig:Mb_GaPPP_PPP_single}]
  {\includegraphics[scale=0.44]{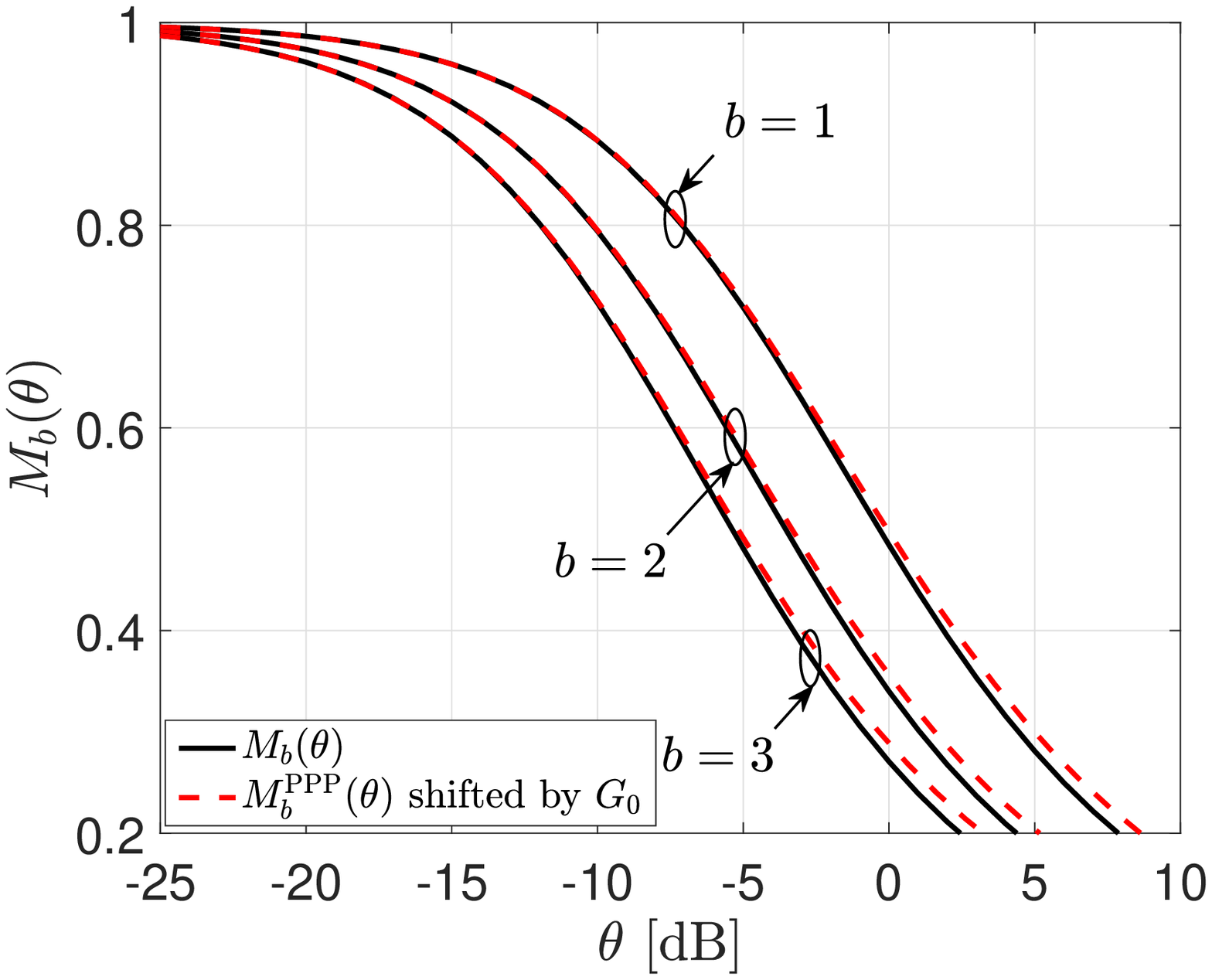}}
   \subfloat[MCP, $G_{0}= -5.1702$ dB\label{fig:Mb_MCP_PPP_single}]
  {\includegraphics[scale=0.44]{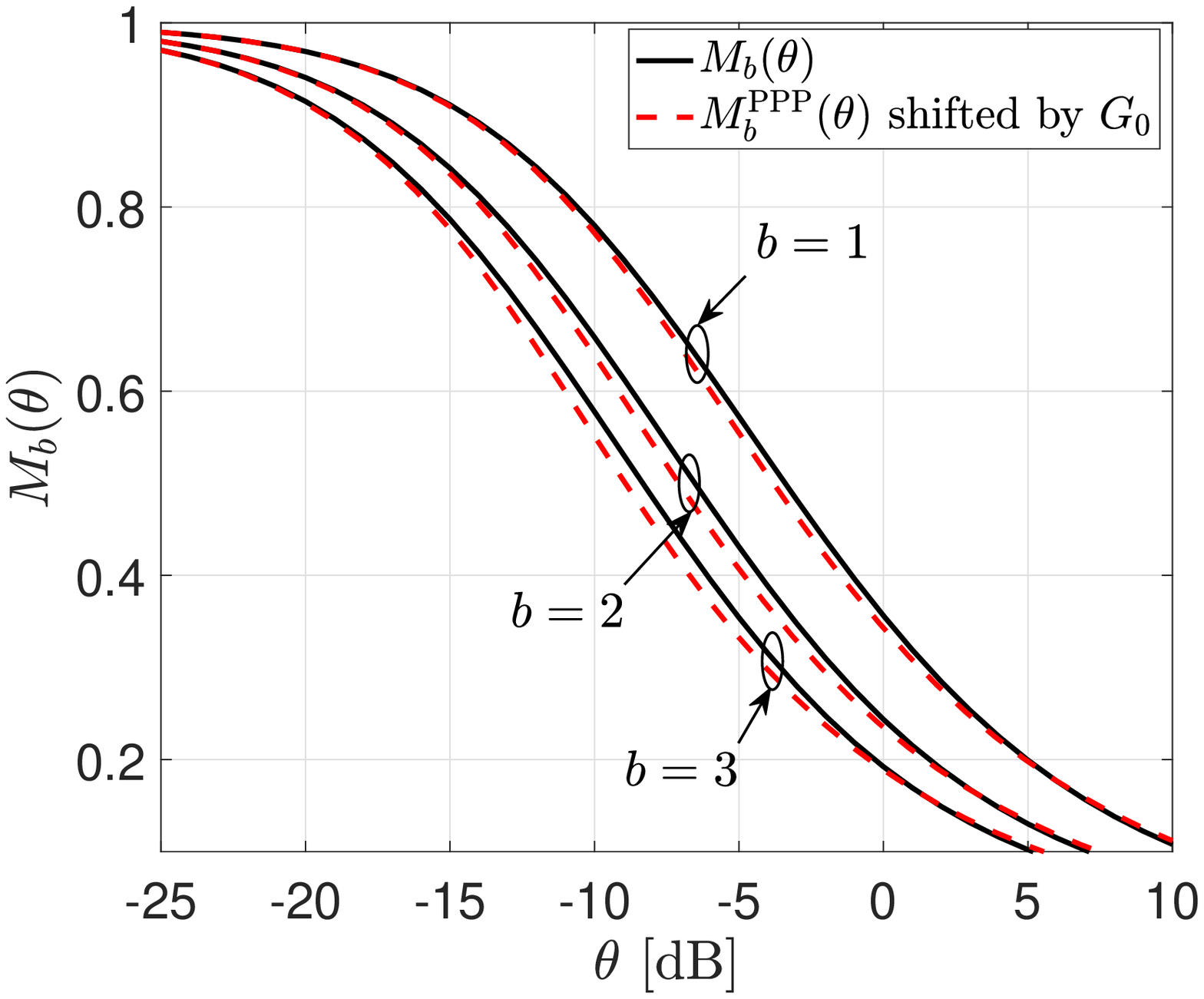}}
\caption{Approximation of $M_b(\theta)$ for four stationary and ergodic non-Poisson point processes by $M_b^{\rm PPP}(\theta/G_0)$. The path loss exponent $\alpha$ is $4$.}
\label{fig:Mb_approximations_single}
\end{figure*}
\subsection{Main Result}

The goal here is to show that an approximation of the form 
\begin{align}
\bar{F}(\theta, x) \approx \bar{F}^{\rm{PPP}}(\theta/G_0, x)
\end{align}
is accurate, where $\bar{F}(\theta, x)$ and  $\bar{F}^{\rm{PPP}}(\theta/G_0, x)$ denote the meta distributions for a stationary and ergodic point process model and a PPP model, respectively.

We know that the shift of the SIR threshold $\theta$ by $G_0$ works quite well for the $1$st moment of the conditional link success probability, {\em i.e.}, the standard success probability $p_{\rm s}(\theta)$, and the meta distribution can be exactly calculated using the moments $M_b(\theta)$ as shown in \eqref{eq:exact_meta}. These results give rise to the interesting question how the $b$th moments $M_b(\theta)$ for an arbitrary stationary and ergodic point process model and $M_b^{\rm{PPP}}(\theta)$ are related to each other, as $\theta \to 0$. The following theorem answers it.
\begin{theorem}
\label{thm:Mb_asym}
For any stationary and ergodic point process and $b \in \mathbb{C}$,
\begin{align}
M_b(\theta) \sim M_b^{\rm{PPP}}(\theta/G_0), \quad \theta \to 0.
\end{align}
\end{theorem}
\begin{IEEEproof}
From \cite[(22)]{martin_meta_2016}, for any stationary and ergodic point process model, we have
\begin{align}
M_{b}(\theta) & = \mathbb{E}\prod_{x \in \Phi \setminus\{x_0\}} \frac{1}{(1+ \theta(\|x_0\|/\|x\|)^{\alpha})^{b}}, \quad b \in \mathbb{C} \\
 &\overset{(\mathrm{a})}{\sim} \mathbb{E} \prod_{y\in \mathcal{R}} (1-b\theta y^\alpha), \quad \theta \to 0 \\
& \sim 1-b\theta \Big(\mathbb{E}\sum_{y\in \mathcal{R}} y^\alpha\Big), \quad \theta \to 0\\
 &\overset{(\mathrm{b})}{=} 1-b\theta\,\mathsf{MISR},
 \label{eq:misr_general}
\end{align}
where $\mathcal{R} \triangleq \{x \in \Phi\setminus \{x_0\}\colon \|x_0\| / \|x\|$\} is the relative distance process (RDP)~\cite[Def. 2]{ganti_asymptotics}, $\mathrm{(a)}$ follows by letting $y=\|x_0\| / \|x\|$ and using Taylor series expansion, and $(\mathrm{b})$ follows from the definition of the MISR for the RDP~\cite{ganti_asymptotics}. Using \eqref{eq:G_0} and \eqref{eq:misr_general}, we obtain the desired result.
\end{IEEEproof}
Note that even though all moments are shifted by the same amount $G_0$ asymptotically, this does not imply that the meta distribution is also shifted by that amount. However, we can expect the shifted meta distribution for the PPP to provide a good approximation. Next, we explore by simulation whether this is the case.

\subsection{Simulation Results}
In this subsection, for a single-tier cellular network, using simulations, we verify the accuracy of approximating the meta distribution for stationary and ergodic point processes by shifting the meta distribution for the PPP by $G_0$.

\textit{Simulation setup}: We perform simulations over a square region $[-500, 500]^2$. Unless otherwise mentioned, we assume the following simulation parameters pertaining to specific point processes:
\begin{itemize}
\item \textit{PTL}: The perturbation $R_{\rm pert}$ is $0.5\eta$. 
\item \textit{GaPPP}: The probability $p$ that a cluster contains one point is $0.5$. Hence the probability that a cluster contains two points is $1 -p = 0.5$. The density of the parent point process is $1/15$. In a two-point cluster, the distance between those two points is $u = 1$. 
\item \textit{MCP}: The density of the parent point process $\lambda_{\rm{p}}$ is $0.01$. The mean number of points $\bar{c}$ in a cluster is $10$. The radius of the disk $r_{\rm{c}}$ around a parent point over which the associated cluster points are distributed is $4$.
\end{itemize}  
We average over $5 \times 10^5$ realizations of the point process.

\begin{figure}
\centering
\includegraphics[width=9cm,height=7.3cm]{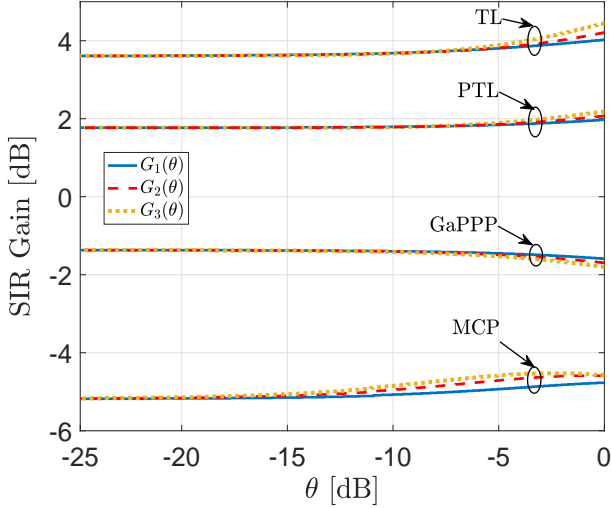}
\caption{The SIR gains $G_{1}(\theta)$, $G_{2}(\theta)$, and $G_{3}(\theta)$ corresponding to the moments $M_1(\theta)$, $M_2(\theta)$, and $M_3(\theta)$, respectively, for the path loss exponent $\alpha = 4$. The asymptotic gain $G_{0}$ is $3.6099$ dB for the TL, $1.8343$ dB for the PTL, $-1.3768$ dB for the GaPPP, and $-5.1702$ dB for the MCP.
}
\label{fig:G1_G2_G3}
\end{figure}

\begin{figure}
\centerline{\epsfig{file=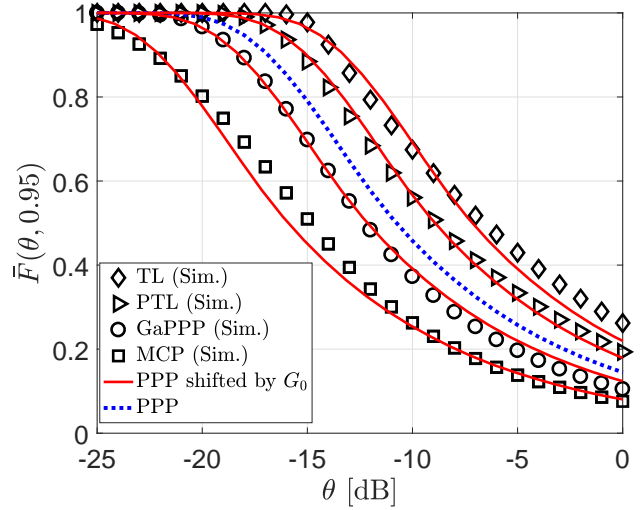,width=9cm,height=7.3cm}}
\caption{The meta distribution $\bar{F}(\theta, x)$ for the PPP, the TL, the PTL, the GaPPP, and the MCP against the SIR threshold $\theta$ for the path loss exponent $\alpha = 4$ and the reliability threshold $x = 0.95$. The asymptotic gain $G_{0}$ is $3.6099$ dB for the TL, $1.8343$ dB for the PTL, $-1.3768$ dB for the GaPPP, and $-5.1702$ dB for the MCP.}
\label{fig:meta_comp_G0}
\end{figure}

\begin{figure}
\centerline{\epsfig{file=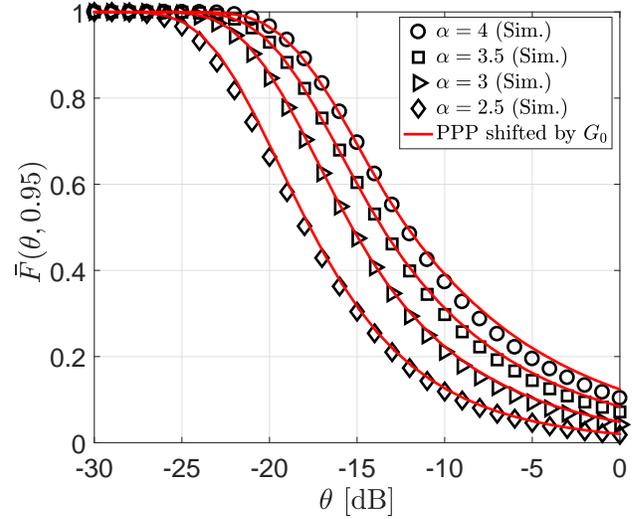,width=9cm,height=7.3cm}}
\caption{The meta distribution $\bar{F}(\theta, x)$ for the GaPPP against the SIR threshold $\theta$ for different values of the path loss exponent and the reliability threshold $x = 0.95$. We have $G_{0} = -1.3768$ dB for $\alpha = 4$, $G_{0} = -1.3423$ dB for $\alpha = 3.5$, $G_{0} = -1.2558$ dB for $\alpha = 3$, and $G_{0} = -0.8661$ dB for $\alpha = 2.5$.}
\label{fig:meta_gpp_diff_alpha}
\end{figure}

\begin{figure}
\centerline{\epsfig{file=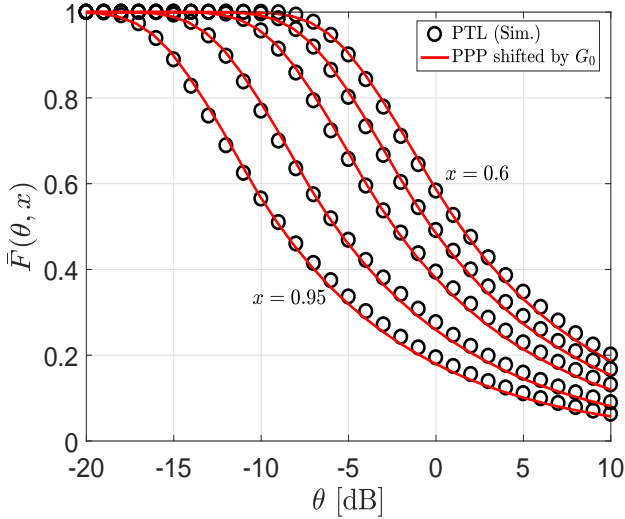,width=9cm,height=7.3cm}}
\caption{The meta distribution $\bar{F}(\theta, x)$ for the PTL against the SIR threshold $\theta$ for different values of the reliability threshold $x$ and the path loss exponent $\alpha = 4$. The asymptotic gain $G_{0}$ for the PTL is $1.8343$ dB. The curves are for $x = 0.6, 0.7, 0.8, 0.9, 0.95$ (from top to bottom).}
\label{fig:single_tier_pert_tri_diff_x_alpha_4}
\end{figure}

Fig.~\ref{fig:Mb_approximations_single} shows that the $b$th moment of the conditional link success probability for a stationary and ergodic non-Poisson point process model can be approximately obtained by shifting that for the PPP model by the asymptotic SIR gain $G_{0}$ for the path loss exponent $\alpha = 4$. The approximation becomes extremely accurate as the SIR threshold $\theta$ becomes small, as expected from Thm.~\ref{thm:Mb_asym}.

Given that the ASAPPP method aims at approximating the first moment $M_1(\theta)$ of the conditional link success probability $P_{\rm{s}}(\theta)$ and the meta distribution can be obtained using the moments (see \eqref{eq:exact_meta}), we numerically calculate the  deployment gain $G_{b}(\theta)$ with respect to the $b$th moment $M_b(\theta)$ for an arbitrary $\theta$. The gain $G_b(\theta)$ is the ratio (gap if measured in dB) $\theta'/\theta$, where $\theta'$ is given by $M_b(\theta') = M_b^{\rm{PPP}}(\theta)$. The moment $M_b^{\rm{PPP}}(\theta)$ is given by~\eqref{eq:Mb_PPP}. For different values of $\theta$, Fig.~\ref{fig:G1_G2_G3} plots the gains $G_{1}(\theta), G_{2}(\theta)$, and $G_{3}(\theta)$ with $\alpha = 4$. Recall that the asymptotic deployment gain $G_0 = \displaystyle \lim_{\theta \to 0} G_1(\theta)$ corresponds to $M_1(\theta)$ as $\theta \to 0$.
We observe from Fig.~\ref{fig:G1_G2_G3} that the asymptotic gain $G_0$ is a good approximation of $G_1(\theta)$, $G_2(\theta)$, and $G_3(\theta)$ for all values of $\theta$ and that the asymptotic value is essentially reached at $\theta=-15$ dB in all cases.

For a single-tier cellular network, Fig.~\ref{fig:meta_comp_G0} plots the meta distribution values for $x = 0.95$ against different SIR thresholds $\theta$ for the PPP, the TL, the PTL, the GaPPP, and the MCP at $\alpha = 4$. We observe that the meta distribution for the TL, the PTL, the GaPPP, and the MCP can be obtained approximately by simply applying a horizontal shift of $G_0$ (in dB) to the meta distribution for the PPP, {\em i.e.}, the AMAPPP can indeed be used to approximately calculate the meta distribution for a non-Poisson network. Especially, for the values of practical interest where a high fraction of users meet the target reliability ({\em e.g.,} the $5\%$ user performance), the approximation is quite accurate. This is quite remarkable given that we use the asymptotic SIR gain $G_0$ as $\theta \to 0$ corresponding to the mean of the distribution to approximate the distribution itself.  

Figs.~\ref{fig:meta_gpp_diff_alpha} and \ref{fig:single_tier_pert_tri_diff_x_alpha_4} confirm the effectiveness of the AMAPPP method for different values of the path loss exponents $\alpha$ and the reliability thresholds $x$, respectively.

\begin{figure}
\centering
\includegraphics[width=9cm,height=7.3cm]{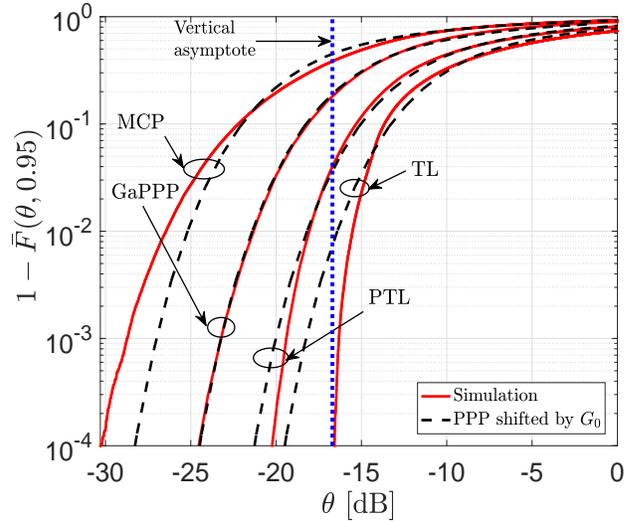}
\caption{The meta distribution $1-\bar{F}(\theta, x)$ against the SIR threshold $\theta$ for the path loss exponent $\alpha = 4$ and the reliability threshold $x = 0.95$. The asymptotic gain $G_{0}$ is $3.6099$ dB for the TL, $1.8343$ dB for the PTL, $-1.3768$ dB for the GaPPP, and $-5.1702$ dB for the MCP. For small $\theta$, the AMAPPP approximation is optimistic for the MCP in that the shifted curve of the PPP is below that of the simulation curve, while the approximation is pessimistic for the regular point processes.}
\label{fig:low_theta_comp}
\end{figure}

\begin{figure}
\centering
\includegraphics[width=9cm,height=7.3cm]{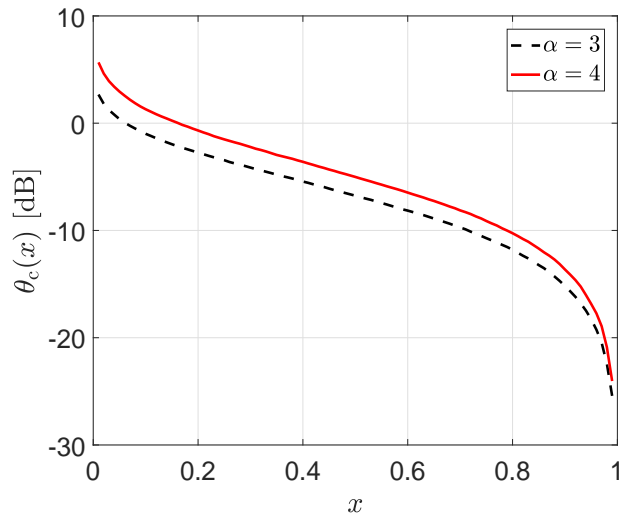}
\caption{The critical SIR threshold $\theta_{\rm c}$ against the reliability threshold $x$ for the triangular lattice.}
\label{fig:crit_theta_x}
\end{figure}

In Fig.~\ref{fig:low_theta_comp}, to check the accuracy of the AMAPPP method for the entire meta distribution, we take a closer look at the regime where the SIR threshold $\theta$ is small. For the GaPPP, the approximation is extremely accurate even for small values of $\theta$, which confirms that the AMAPPP method is accurate for the entire meta distribution for the GaPPP. For the MCP, there is a gap between the simulation and the approximation (the shifted PPP) because the asymptotics {\em kick in} slowly compared to that for the GaPPP.
This can be confirmed from Fig.~\ref{fig:G1_G2_G3}, which shows that, for the MCP, the gains $G_1(\theta)$, $G_2(\theta)$, and $G_3(\theta)$ corresponding to $M_1(\theta)$, $M_2(\theta)$, and $M_3(\theta)$, respectively, converge to $G_0$ more slowly than those for the GaPPP as $\theta$ becomes small. 

For the stationary triangular lattice, Fig.~\ref{fig:low_theta_comp} highlights an interesting case; at $\theta = -16.68$ dB, the value of $1-\bar{F}(\theta, x)$ drops to zero because all users achieve the target reliability $x = 0.95$ for $\theta < -16.68$ dB. For all $x$, such a threshold can be calculated by shifting the lattice such that the typical user sits at a Voronoi vertex for the triangular lattice, which is the worst-case scenario for the triangular lattice since there are 3 nearest base stations to the user. For all lattices, there exists such a vertical asymptote, and thus the approximation by shifting $G_0$ breaks down as $\theta$ approaches that threshold. However, for values of $\theta$ for which the $5\%$ user achieves $95\%$ reliability, the approximation is tight. Such a worst-case scenario could also occur in the case of the perturbed triangular lattice if the perturbation radius $R_{\rm pert}$ is small enough. In this case, the critical $\theta$ is smaller compared to that for the stationary triangular lattice. For example, at $R_{\rm pert} = 0.5\eta$, the approximation starts to break down at a smaller value of $\theta$ compared to that for the stationary triangular lattice. 

Fig.~\ref{fig:crit_theta_x} shows that the critical $\theta$, denoted by $\theta_{\rm c}(x)$, depends on the target reliability $x$---it decreases with an increase in $x$ because, to meet a higher target reliability, the SIR threshold $\theta$ at the worst-case user needs to be lowered. Such a dependency of $\theta_{\rm c}(x)$ on the target reliability illustrates the rate-reliability trade-off.\footnote{The rate is a function of the SIR threshold.} Fig.~\ref{fig:crit_theta_x} also provides insight into the support of the probability density function (pdf) of the conditional link success probability $P_{\rm s}(\theta)$. For each $x>0$, there is a $\theta_{\rm c}(x)$ such that the support is reduced to $[x,1]$ at $\theta = \theta_{\rm c}(x)$. For $\theta < \theta_{\rm c}(x)$, the support gradually reduces to $\{1\}$.

\section{Heterogeneous Cellular Networks}
\label{sec:AMAPPP_HCN}
\subsection{Per-tier AMAPPP Approximation}
\label{sec:per_tier_AMAPPP}
We consider a $K$-tier general cellular network where the $i$th tier is modeled by a stationary and ergodic point process $\Phi_i$ of density $\lambda_i$ and each tier is independent of the other tiers. For this setup, the following theorem provides an approximation of $M_b(\theta)$ for general HCNs.
\begin{theorem}
\label{thm:Mb_asym_HCN}
Let 
\begin{align}
\label{eq:approx_Mb}
\hat{M}_b(\theta) &= \sum_{k \in [K]} \int_{0}^{\infty}  \exp\Bigg(-s F(b,\delta_k,\theta/G_k)\nonumber\\
& - \sum_{i \in [K]^{!}}\rho_{ik}s^{\frac{\alpha_k}{\alpha_i}} F(b, \delta_i,\theta)\Bigg){\rm d}s,
\end{align}
where $F(b, \delta, \theta) \triangleq \:_2F_1(b, -\delta; 1-\delta; -\theta)$, $\delta_i \triangleq 2/\alpha_i$, $G_k$ is the asymptotic SIR gain corresponding to the point process that models the BS deployment in the $k$th tier, and $\rho_{ik} = \frac{\lambda_i \pi P_{ik}^{\delta_i}}{(\lambda_k \pi)^{\alpha_k/\alpha_i}}$ with $P_{ik} \triangleq P_i/P_k$.

For $K$-tier general HCNs where the typical cellular user is connected to the BS that results in the strongest average received power, the $b$th moment $M_b(\theta)$ of the conditional link success probability is approximated as
\begin{align}
M_b(\theta) \approx \hat{M}_b(\theta).
\end{align}

\end{theorem}
\begin{IEEEproof}
See Appendix~\ref{app:Mb_asym_HCN}.
\end{IEEEproof}

In this approach, the $b$th moment of the conditional link success probability corresponding to each tier modeled by a non-Poisson point process is approximated by that of the PPP by shifting the SIR threshold by the MISR-based gain $G_k$, which can be further used to calculate the approximate meta distribution using the Gil-Pelaez (GP) theorem or the beta distribution approximation. Hence we call this approach the ``per-tier AMAPPP'' approach. 

In the calculation of the meta distribution using the beta distribution approximation, there are two approximations involved:
\begin{itemize}
\item[1)] The approximate calculation of the first and second moments based on the MISR gain as shown in Thm.~\ref{thm:Mb_asym_HCN},
\item[2)] The inherent approximation resulting from matching only the first and the second moments.
\end{itemize}
Hence we call the combination of the beta distribution approximation with the per-tier AMAPPP approach the ``approximate beta approximation'' (ABA).

When all tiers have the same path loss exponent, \eqref{eq:approx_Mb} reduces to the expression given in the following corollary.
\begin{corollary} 
If $\alpha_1 = \alpha_2 = \dotsc = \alpha_K = \alpha$, we have
\begin{align}
\hat{M}_b(\theta) = \sum_{k \in [K]} \frac{1}{F(b, \delta, \theta/G_k) + \sum_{i\in [K]^{!}}\frac{\lambda_i}{\lambda_k}\left(\frac{P_i}{P_k}\right)^{\delta}F(b,\delta,\theta)}.
\label{eq:same_path_loss}
\end{align}
\end{corollary}
For a $K$-tier HIP cellular network, we have $G_k = 1$. Then when all tiers have the same path loss exponent, \eqref{eq:same_path_loss} simplifies to  
\begin{align}
\hat{M}_{b}(\theta) = 1/F(b,\delta, \theta),
\label{eq:same_path_loss_HIP}
\end{align}
as stated in~\cite[Cor. 1]{Yuanjie_HCN}. Note that \eqref{eq:same_path_loss_HIP} is the exact expression of the $b$th moment of the conditional link success probability for the HIP model, {\em i.e.}, $M_b^{\rm HIP}(\theta) = \hat{M}_{b}(\theta)$.

\begin{figure*}
  \centering
  \subfloat[TL/PPP, $G_{\rm eff}= 1.2190$ dB.\label{fig:Mb_TL_PPP}]{\includegraphics[scale=0.42]{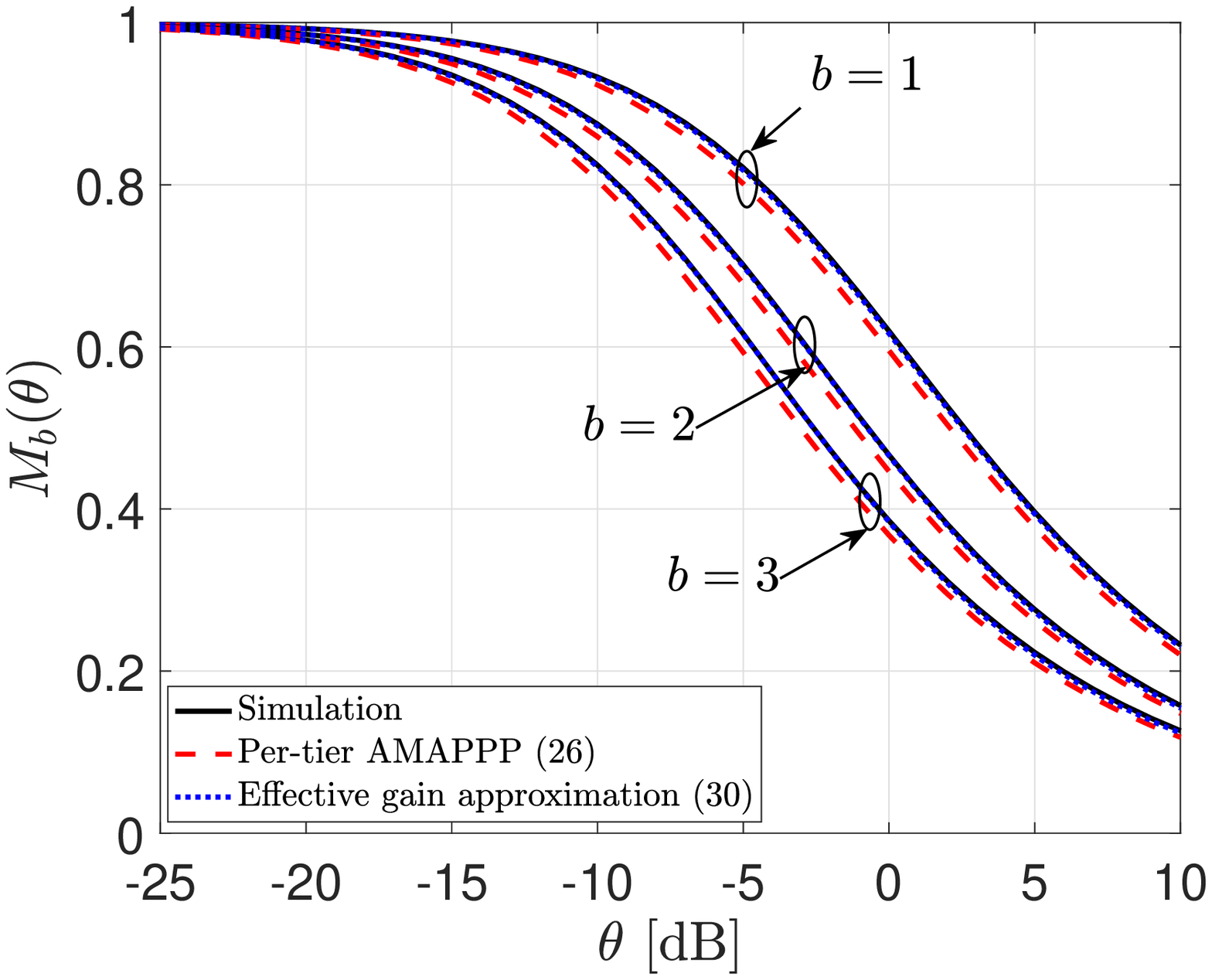}}\hspace{2mm}
  \subfloat[PTL/PPP, $G_{\rm eff}= 0.5361$ dB.\label{fig:Mb_PTL_PPP}]{\includegraphics[scale=0.42]{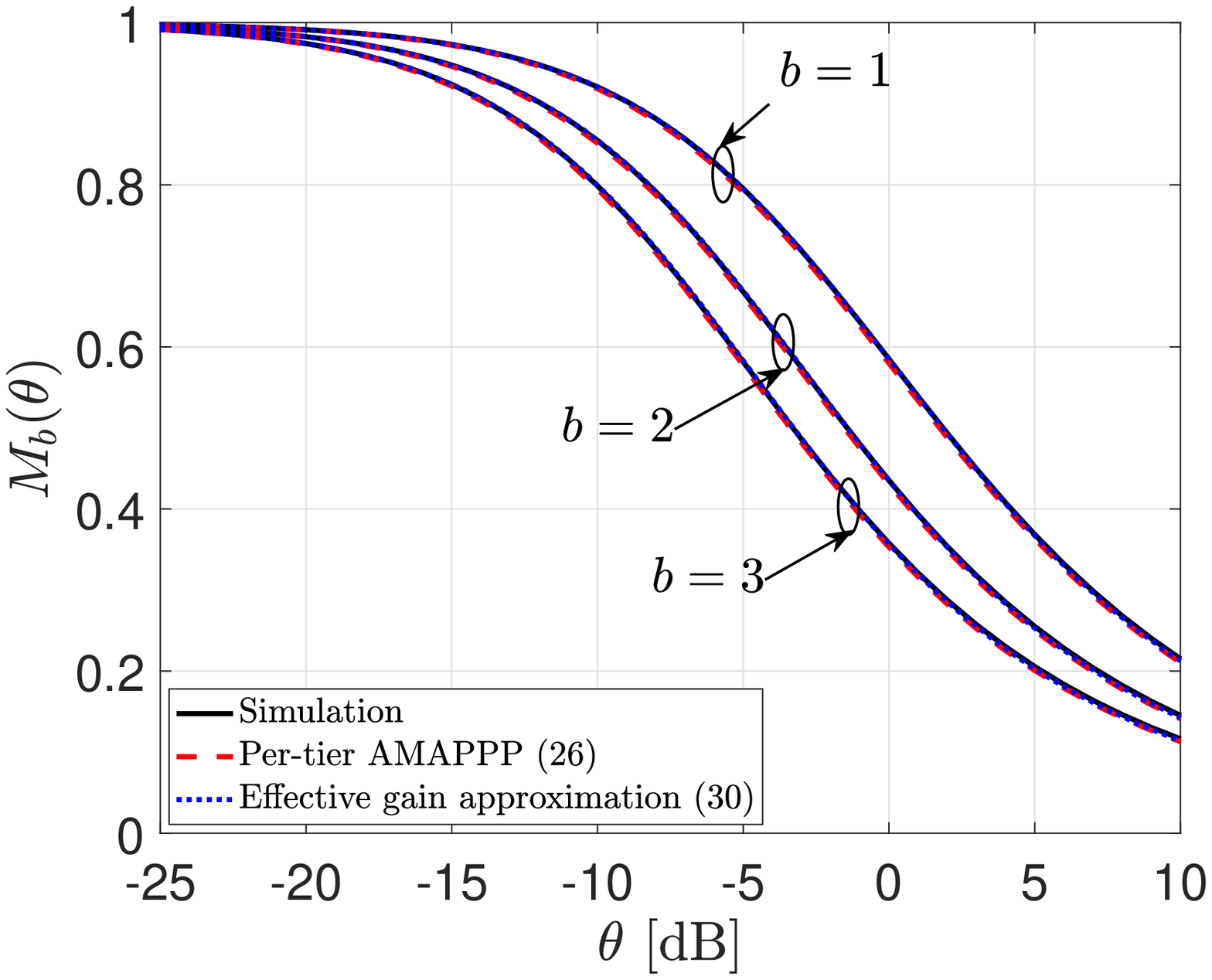}}\\
   \subfloat[GaPPP/PPP, $G_{\rm eff}= -0.3064$ dB.\label{fig:Mb_GaPPP_PPP}]
  {\includegraphics[scale=0.42]{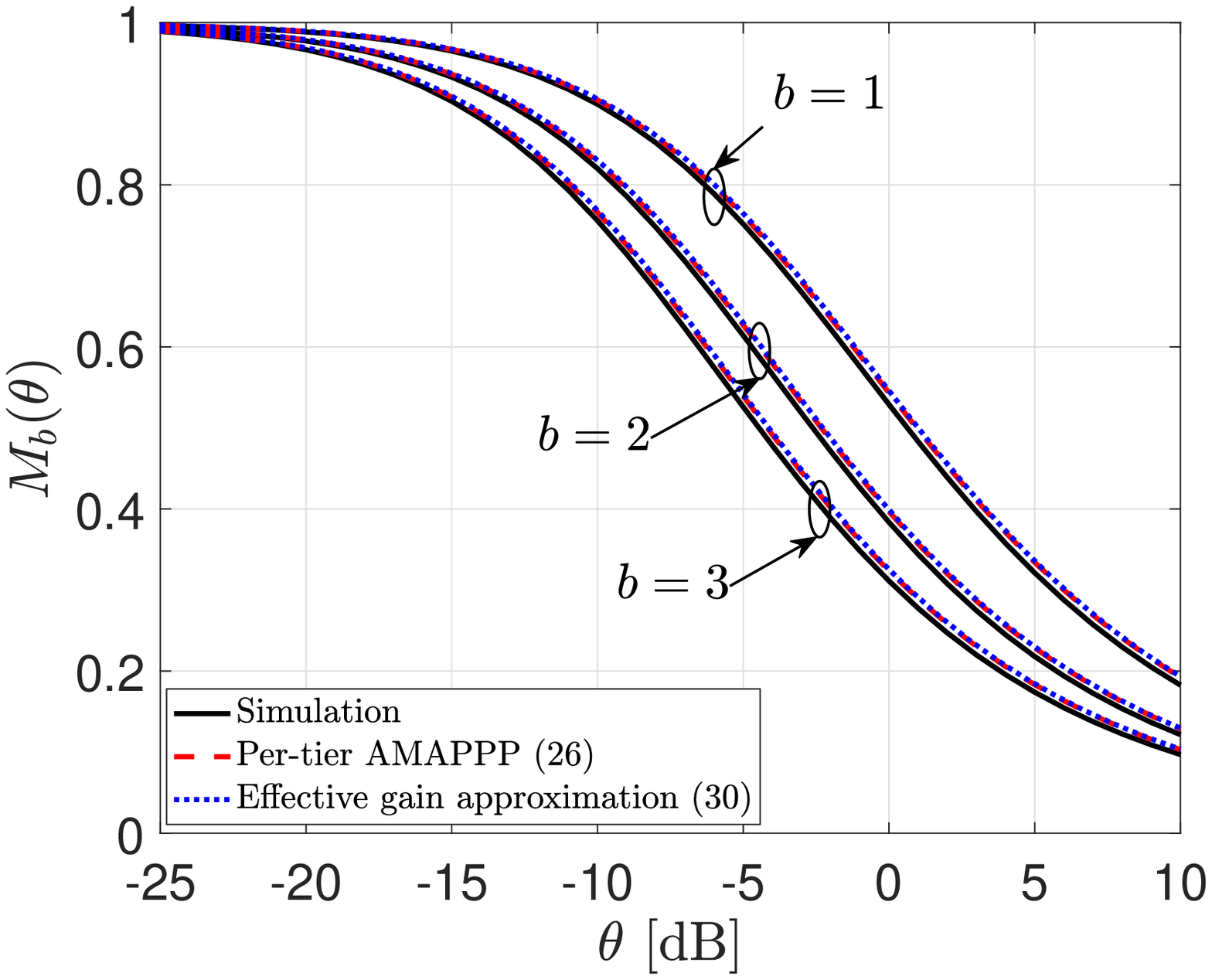}}\hspace{2mm}
   \subfloat[MCP/PPP, $G_{\rm eff}= -0.8301$ dB.\label{fig:Mb_MCP_PPP}]
  {\includegraphics[scale=0.42]{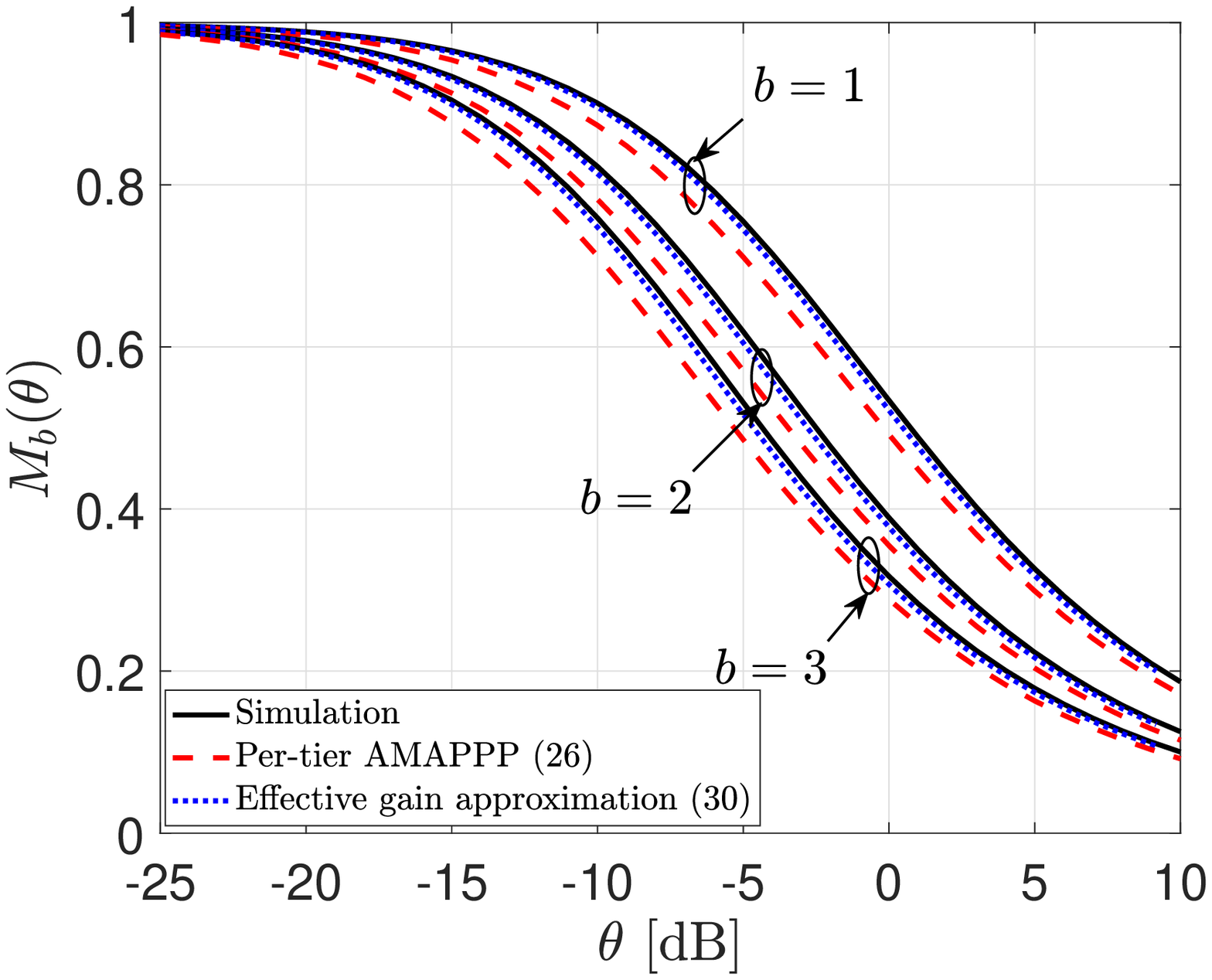}}
\caption{The per-tier AMAPPP approximation and the effective-gain approximation of $M_b(\theta)$ for a ``Non-Poisson/PPP'' deployment (a two-tier general HCN). The path loss exponent is $\alpha = 4$.}
\label{fig:Mb_HCN_approximations}
\end{figure*}
\subsection{Effective Gain Approximation}
In the per-tier AMAPPP method, a non-Poisson tier is approximated by the PPP using the MISR-based gain. In this subsection, we provide an approximation where we directly shift the meta distribution for the $K$-tier HIP model by the {\em effective gain} to approximate the meta distribution for the general $K$-tier cellular network. This effective gain was introduced in \cite{mh_general_asappp} for the standard success probability. The following theorem calculates the effective gain corresponding to $M_{b}$.
\begin{theorem}
\label{thm:G_eff}
When $\alpha_1 = \alpha_2 = \dotsc = \alpha_K = \alpha$, for any $b > 0$, the approximate $b$th moment $\hat{M}_b(\theta)$ for a general K-tier cellular network is related to the $b$th moment $M_b^{\rm HIP}$ for the HIP model as
\begin{align*}
\hat{M}_b(\theta) \leq M_b^{\rm HIP}\left(\frac{\theta}{G_{\rm eff}}\right),
\end{align*}
where 
\begin{align}
G_{\rm eff} &\triangleq \sum_{k \in [K]}w_k(w_k G_k + (1-w_k)) \nonumber \\
&= 1 + \sum_{k \in [K]} w_k^2(G_k - 1),
\label{eq:g_eff}
\end{align}
with $w_k \triangleq \frac{\lambda_k P_k^{\delta}}{\sum_{i \in [K]}\lambda_i P_i^{\delta}}$. For $b < 0$, we need to replace `$\leq$' by `$\geq$'.
\end{theorem}
\begin{IEEEproof}
See Appendix~\ref{app:G_eff}.
\end{IEEEproof}
Using Thm.~\ref{thm:G_eff}, we obtain an another approximation of $M_b(\theta)$ for a stationary and ergodic point process as
\begin{align}
M_b(\theta) \approx M_b^{\rm HIP}\left(\frac{\theta}{G_{\rm eff}}\right).
\end{align}
The effective gain $G_{\rm eff}$ corresponds to the overall SIR gain of HCNs which can be obtained from the MISR-based gains of the individual tiers of the HCNs. Hence, similar to approximating the meta distribution for a stationary and ergodic non-Poisson tier by shifting that for the PPP by the MISR-based gain, the meta distribution for general HCNs can be approximated by directly shifting that for the HIP model by the overall SIR gain, {\em i.e.,}
\begin{align}
\bar{F}(\theta, x) \approx \bar{F}^{\rm HIP}(\theta/G_{\rm eff}, x).
\label{eq:G_eff_approx}
\end{align}
We call the approximation in \eqref{eq:G_eff_approx} the ``effective gain approximation.''

\begin{figure*}
  \centering
  \subfloat[TL/PPP, $G_{\rm eff}= 1.1951$ dB for $\alpha = 3$, $G_{\rm eff}= 1.2190$ dB for $\alpha = 4$.\label{fig:TL_PPP}]{\includegraphics[scale=0.43]{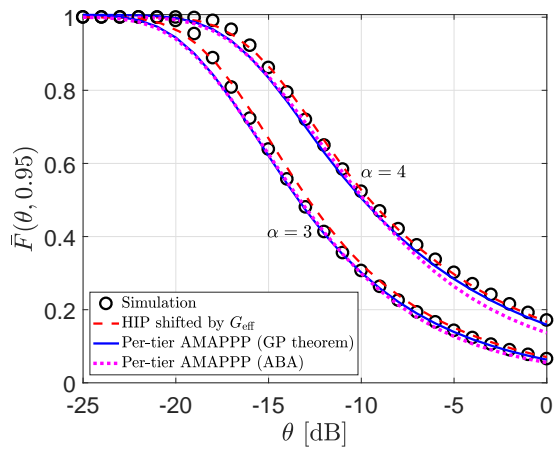}}\hspace{2mm}
  \subfloat[PTL/PPP, $G_{\rm eff}= 0.5491$ dB for $\alpha = 3$, $G_{\rm eff}= 0.5361$ dB for $\alpha = 4$.\label{fig:PTL_PPP}]{\includegraphics[scale=0.43]{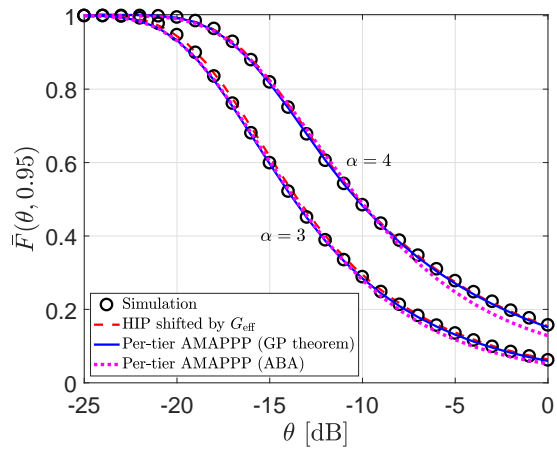}}\\
   \subfloat[GaPPP/PPP, $G_{\rm eff}= -0.2819$ dB for $\alpha = 3$, $G_{\rm eff}= -0.3064$ dB for $\alpha = 4$.\label{fig:GaPPP_PPP}]
  {\includegraphics[scale=0.43]{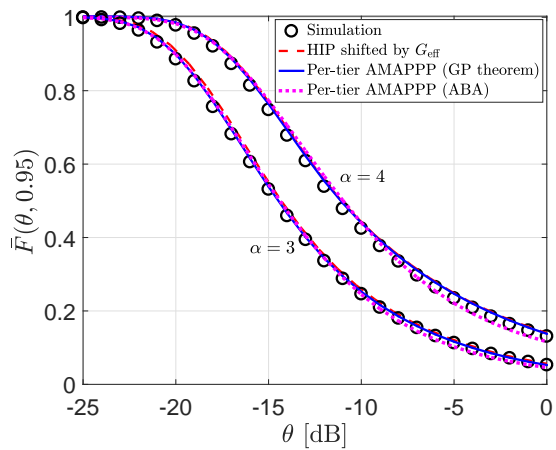}}\hspace{2mm}
   \subfloat[MCP/PPP, $G_{\rm eff}= -0.8511$ dB for $\alpha = 3$, $G_{\rm eff}= -0.8301$ dB for $\alpha = 4$.\label{fig:MCP_PPP}]
  {\includegraphics[scale=0.43]{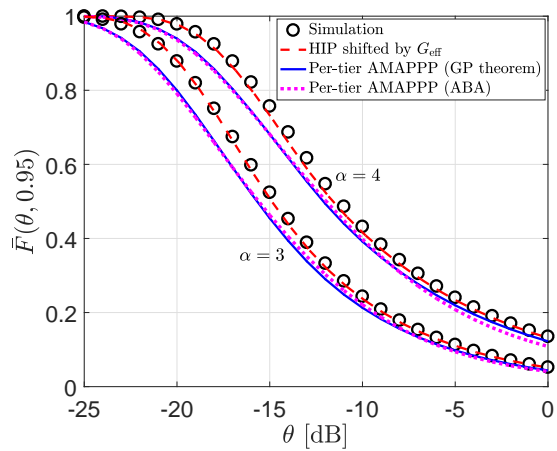}}
\caption{The effective-gain approximation, the per-tier AMAPPP approximation (GP theorem), the approximate beta approximation (ABA) of the meta distribution $\bar{F}(\theta, x)$ of a ``Non-Poisson/PPP'' deployment (a two-tier general HCN) against $\theta$.}
\label{fig:HCN_approximations}
\end{figure*}

\subsection{Results}
The simulation parameters are the same as those for a single-tier cellular network. Unless otherwise mentioned, each tier has a BS density of $0.1$. We assume $\alpha_1= \alpha_2 = \dotsc = \alpha_K = \alpha$.

For a $2$-tier HCN denoted by {\em first tier/second tier}, Fig.~\ref{fig:Mb_HCN_approximations} shows that the per-tier AMPPPP approximation and the effective gain approximation closely approximate the $b$th moment $M_b(\theta)$ of the conditional link success probability. As proved in Thm.~\ref{thm:G_eff}, the effective gain approximation provides an upper bound on the per-tier approximation. Fig.~\ref{fig:Mb_HCN_approximations} also shows that the gap between the simulation and the per-tier AMAPPP approximation is larger for the TL compared to that for the PTL since the former is more regular than the latter, and thus the approximation of  the interference by that of the PPP is less accurate. Similarly, the gap between the simulation and the per-tier AMAPPP approximation is smaller for the GaPPP compared to that for the MCP because the latter is more clustered than the former.

\begin{figure}
\centerline{\epsfig{file=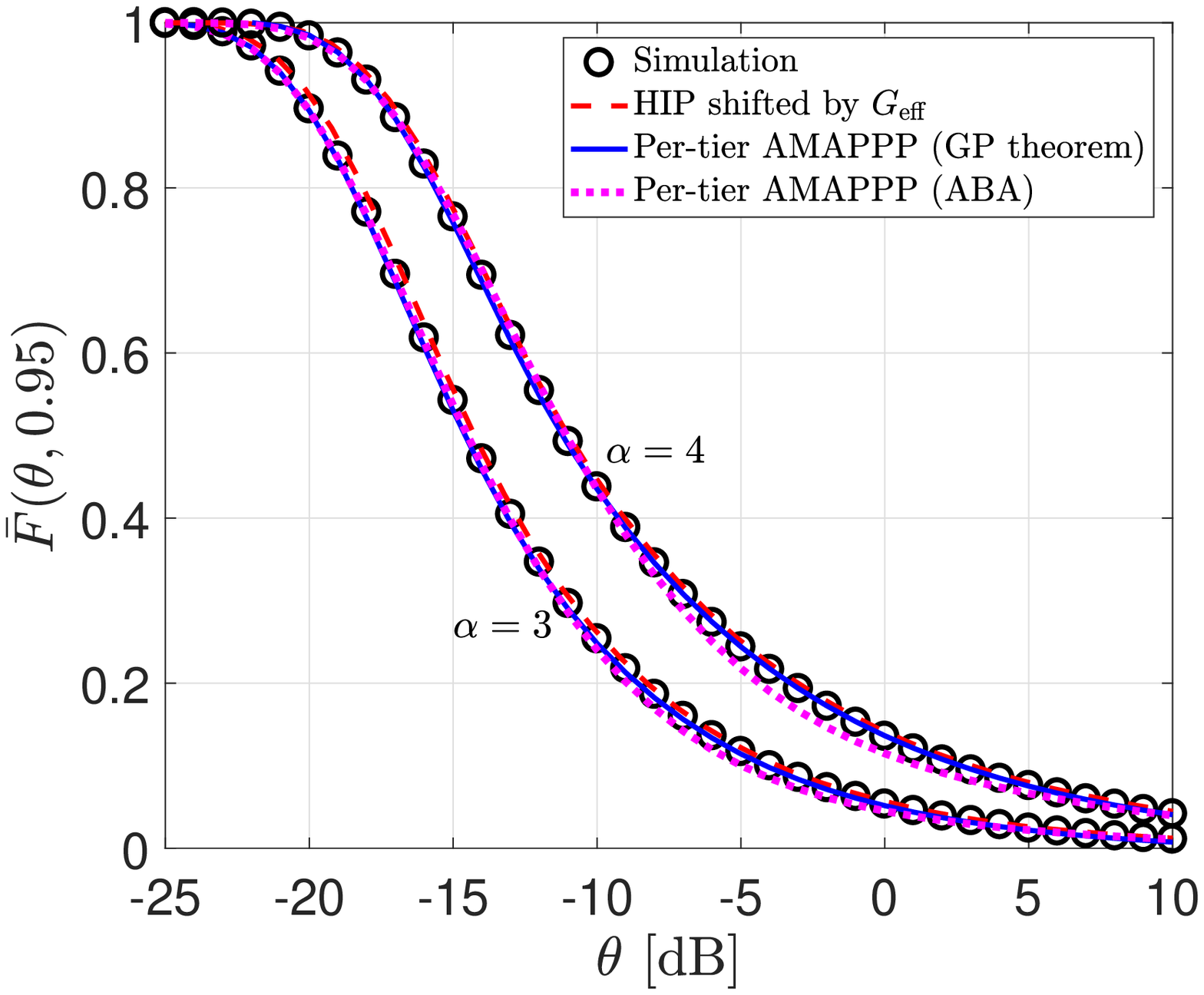,width=9cm,height=7.3cm}}
\caption{The meta distribution $\bar{F}(\theta, x)$ for a 2-tier cellular network GaPPP/PPP against the SIR threshold $\theta$ for the reliability threshold $x = 0.95$ and the path loss exponent $\alpha = 4$. The density of the GaPPP is $0.2$, while the density of the PPP is $0.1$. For $\alpha = 3$, $G_{\rm eff} = -0.2226$ dB and for $\alpha = 4$, $G_{\rm eff} = -0.2287$ dB.}
\label{fig:amappp_diff_dens}
\end{figure}

\begin{figure}
\centerline{\epsfig{file=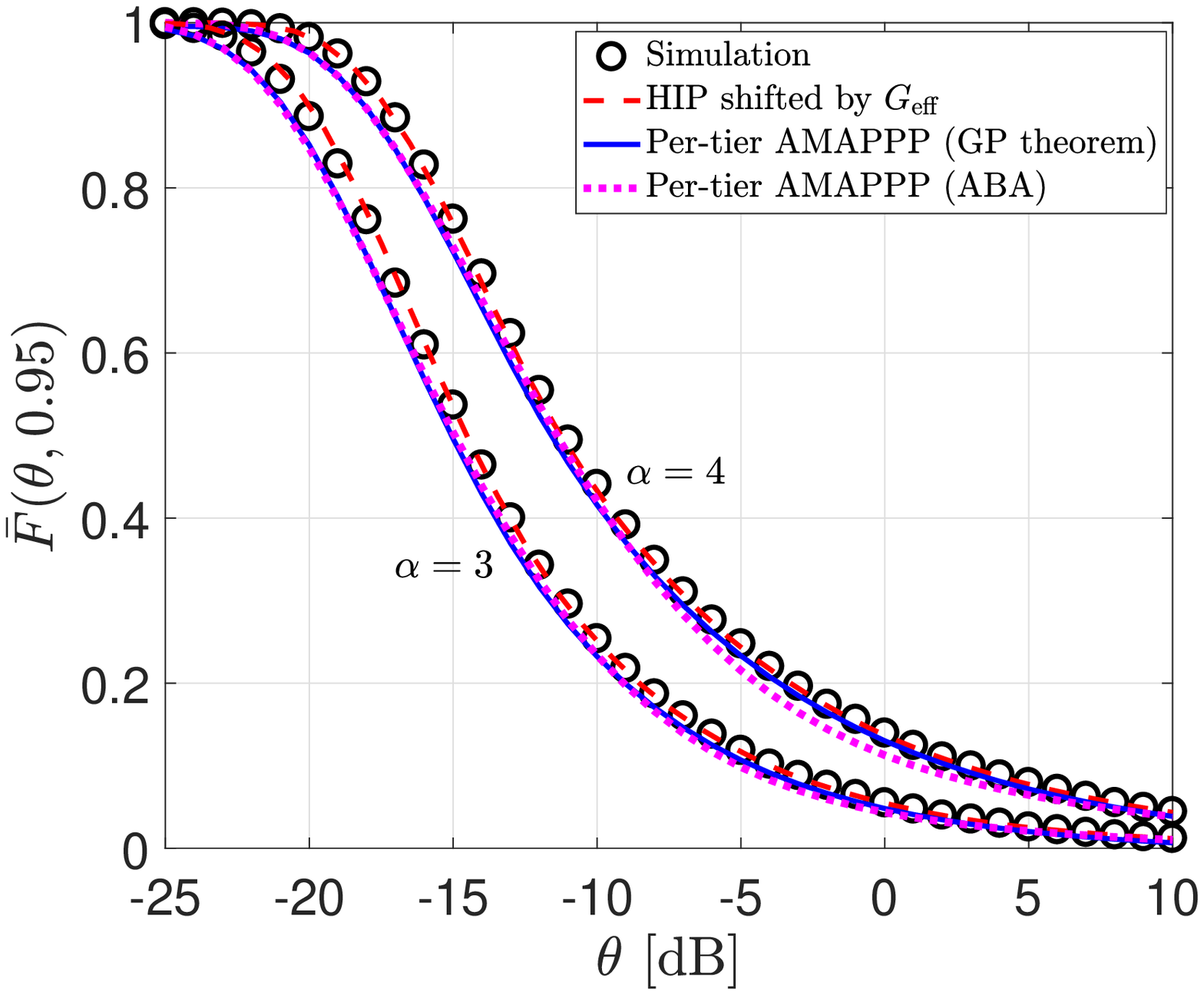,width=9cm,height=7.3cm}}
\caption{The meta distribution $\bar{F}(\theta, x)$ for a $3$-tier cellular network GaPPP/MCP/PPP against the SIR threshold $\theta$ for the reliability threshold $x = 0.95$ and the path loss exponent $\alpha = 4$. For $\alpha = 3$, $G_{\rm eff}= -0.4910$ dB and for $\alpha = 4$, $G_{\rm eff}= -0.4959$ dB.}
\label{fig:meta_gpp_mcp_ppp_alpha_3_4}
\end{figure}

For the path loss exponents of $\alpha = 3, 4$, Fig.~\ref{fig:HCN_approximations} shows the meta distribution for a $2$-tier HCN. The BSs of the first tier form a stationary and ergodic non-Poisson point process and those of the second tier a PPP. The effective gain approximation and the per-tier approximation both work quite well over a wide range of $\theta$. Especially, the approximate beta approximation (ABA) method is remarkably accurate given that it involves two approximations. Fig.~\ref{fig:amappp_diff_dens} shows the accuracy of the AMAPPP approximation when each tier of a multi-tier cellular network is modeled by a stationary and ergodic point process of different densities.
Fig.~\ref{fig:meta_gpp_mcp_ppp_alpha_3_4} verifies the accuracy of both approximations for a $3$-tier general HCN. 
\begin{figure}
\centerline{\epsfig{file=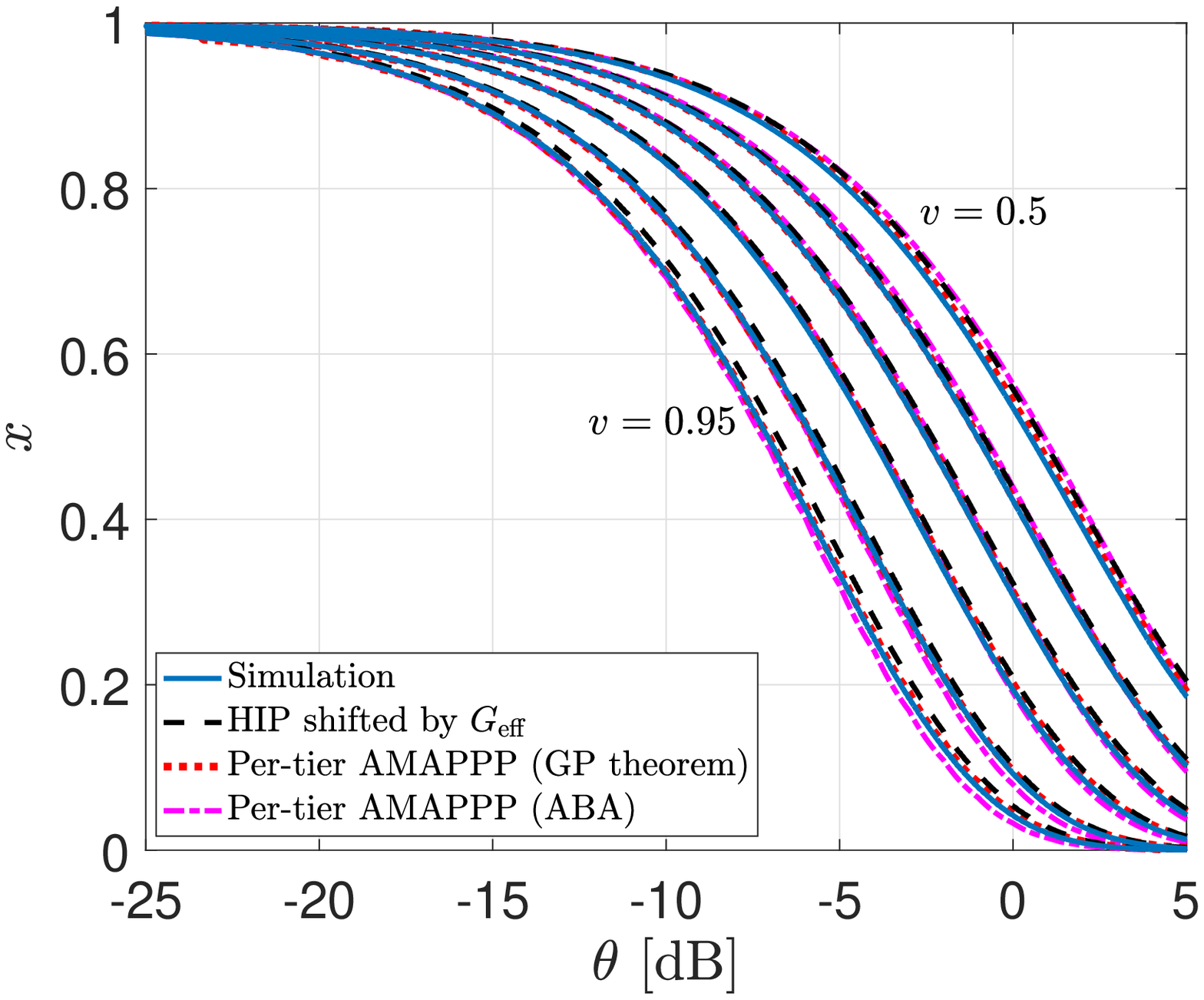,width=9cm,height=7.3cm}}
\caption{Contour plot of the meta distribution $\bar{F}(\theta, x)$ for the GaPPP/PPP cellular network for the path loss exponent $\alpha = 4$. The effective gain is $G_{\rm eff} = -0.3064$ dB. The values at the curves are $\bar{F}(\theta, x) = v = 0.95, 0.9, 0.8, 0.7, 0.6,$ and $0.5$ (from bottom to top).}
\label{fig:contour_meta_gpp_ppp_sing_alpha_4}
\end{figure}

\begin{figure}
\centerline{\epsfig{file=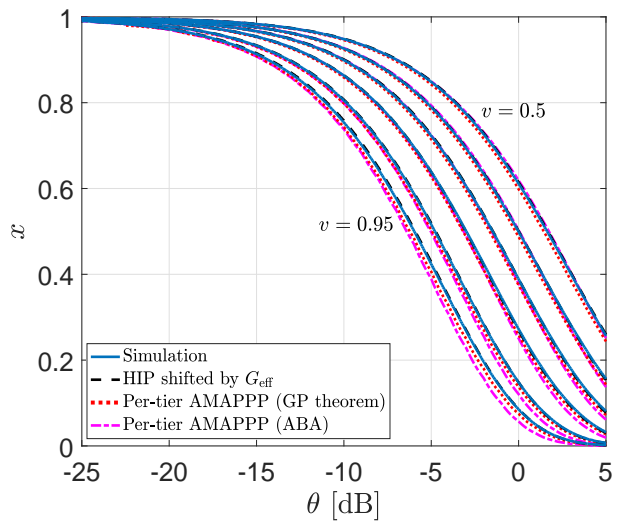,width=9cm,height=7.3cm}}
\caption{Contour plot of the meta distribution $\bar{F}(\theta, x)$ for the PTL/PPP cellular network for the path loss exponent $\alpha = 4$. The effective gain is $G_{\rm eff} = 0.5361$ dB. The values at the curves are $\bar{F}(\theta, x) = v = 0.95, 0.9, 0.8, 0.7, 0.6,$ and $0.5$ (from bottom to top).}
\label{fig:contour_meta_pert_tri_ppp_alpha_4_theta_x}
\end{figure}

Figs.~\ref{fig:contour_meta_gpp_ppp_sing_alpha_4} and \ref{fig:contour_meta_pert_tri_ppp_alpha_4_theta_x} show the contour plots for $2$-tier HCNs. These contour plots illustrate the trade-off between the SIR threshold $\theta$ and the reliability threshold $x$. The contours provide the possible pairs ($\theta, x$) that a fraction $v$ of users achieves. For example, in Fig.~\ref{fig:contour_meta_gpp_ppp_sing_alpha_4}, the curve corresponding to $v = 0.95$ shows that $95\%$ users achieve an SIR of $-10$ dB with probability $0.69$ and an SIR of $-5$ dB with probability $0.34$. Furthermore, Figs. \ref{fig:contour_meta_gpp_ppp_sing_alpha_4} and \ref{fig:contour_meta_pert_tri_ppp_alpha_4_theta_x} show that the effective gain approximation and the per-tier AMAPPP method (using both the GP theorem and ABA) work quite well for different values of the fraction $v$ of users, the reliability threshold $x$, and the SIR threshold $\theta$.

\section{Conclusions}
\label{sec:conclusions}
In this paper, we have proposed AMAPPP, a simple and novel approach to approximately obtain the meta distribution for an arbitrary stationary and ergodic point process as well as general HCNs from that for the PPP and the HIP model, respectively. For the $b$th moment $M_b(\theta)$ of the conditional success probability for any stationary and ergodic point process model, we proved that $M_{b}(\theta) \sim 1 - b\theta\, \mathsf{MISR}$, as $\theta \to 0$. Through detailed simulations for the triangular lattice, the perturbed triangular lattice, the Gauss-Poisson point process, and the Mat{\'e}rn cluster process, we have shown that the asymptotic deployment gain $G_0$ of the standard success (coverage) probability can be used to relate the meta distribution to that for the PPP. The approximation of the meta distribution for the triangular lattice by that for the PPP becomes pessimistic in the worst-case scenario, \textit{i.e.}, when the typical cellular user is located such that it has three nearest base stations. For $K$-tier HCNs, the per-tier approach closely approximates the $b$th moment of the conditional link success probability, which can be further used to calculate the approximate meta distribution. The other approach directly calculates the approximate meta distribution for general HCNs from that for the HIP model by simply applying a shift by the effective gain. Overall, given the generality of the model and the fine-grained nature of the meta distribution, the AMAPPP approach works surprisingly well.

There are interesting future directions to our work. It is important to obtain PPP-based approximations for the SIR meta distribution for uplink and general fading models. But one has to be careful in the uplink case since there could be an interferer arbitrarily close to a receiver, which makes the uplink problem intricate. Also there is no work on the SIR meta distribution with general fading models even for the PPP-based models. Hence it is naturally more appropriate to first analyze the SIR meta distribution for the PPP with general fading and then investigate whether the shifting approach works for general cellular networks with general fading models. Another interesting line of research is to consider the association with a base station that offers the maximum instantaneous SIR, instead of the nearest-base station association.

\appendices
\section{Proof of Thm.~\ref{thm:Mb_asym_HCN}}
\label{app:Mb_asym_HCN}
When the typical user $x_0$ connects to a BS in the $k$th tier, the conditional link success probability for the typical user is given as
\begin{align}
P_{\rm s}^{(k)}(\theta) &= \mathbb{P}\Bigg(\frac{P_k h_0 \ell_k(x_0)}{I}  > \theta, x_0 \in \Phi_{k} \mid  \Phi_1, \dotsc, \Phi_K \Bigg) \nonumber \\
& = \mathbb{E}\Bigg[\exp\Bigg(-\theta \frac{I}{P_k  \ell_k(x_0)}\Bigg)\boldsymbol{1}_{x_0 \in \Phi_k}\mid \Phi_1, \dotsc, \Phi_K \Bigg],
\end{align}
where $P_k$ is the transmit power of a BS associated with the $k$th tier, $[K] = \lbrace 1, 2, \dotsc, K \rbrace$, $[K]^{!} = [K]\setminus \{k\}$, and $I = \sum_{x \in \Phi_k^{!}}P_k h_x \ell_k(x) + \sum_{i \in [K]^{!}}\sum_{y \in \Phi_i}P_i h_y \ell_i(y)$. $\boldsymbol{1}$ denotes the indicator function.

Averaging over the fading, it follows that
\begin{align}
P_{\rm s}^{(k)}(\theta) &=  \prod_{x \in \Phi_k^{!}} \frac{1}{1+\frac{\theta \ell_k(x)}{\ell_k(x_0)}} \prod_{i \in [K]^{!}}\prod_{y \in \Phi_i} \frac{1}{1+\frac{\theta P_{ik} \ell_i(y)}{\ell_k(x_0)}}\boldsymbol{1}_{x_0 \in \Phi_k},
\end{align}
where $P_{ik} = P_i/P_k$.
The $b$th moment of $P_{\rm s}^{(k)}$ follows as
\begin{align}
&M_b^{(k)}(\theta) \nonumber \\
&=\mathbb{E}\!\Bigg[ \prod_{x \in \Phi_k^{!}} \!\frac{1}{\left(1+\frac{\theta \ell_k(x)}{\ell_k(x_0)}\right)^{b}} \!\prod_{i \in [K]^{!}}\!\prod_{y \in \Phi_i} \!\frac{1}{\left(1+\frac{\theta P_{ik}\ell_i(y)}{\ell_k(x_0)}\right)^b}\boldsymbol{1}_{x_0 \in \Phi_k}\!\Bigg] \nonumber\\
&\overset{(\mathrm{a})}{\sim} \mathbb{E}\Bigg[\prod_{x \in \Phi^{!{\rm PPP}}_{k}} \frac{1}{\left(1+\frac{\theta\ell_k(x)}{G_k\ell_k(x_0)}\right)^{b}} \nonumber \\
&\times \prod_{i \in [K]^{!}}\prod_{y \in \Phi_i} \frac{1}{\left(1+\frac{\theta P_{ik}\ell_i(y)}{\ell_k(x_0)}\right)^b}\boldsymbol{1}_{x_0 \in \Phi_k^{!{\rm PPP}}}\Bigg] \nonumber \\
&\overset{({\mathrm{b}})}{\approx} \mathbb{E}\Bigg[\prod_{x \in \Phi^{!{\rm PPP}}_{k}} \frac{1}{\left(1+\frac{\theta\ell_k(x)}{G_k\ell_k(x_0)}\right)^{b}} \nonumber \\
& \times \prod_{i \in [K]^{!}}\prod_{y \in \Phi_{i}^{\rm PPP}} \frac{1}{\left(1+\frac{\theta P_{ik}\ell_i(y)}{\ell_k(x_0)}\right)^b} \boldsymbol{1}_{x_0 \in \Phi_k^{!{\rm PPP}}}\Bigg] \nonumber \\
& \overset{(\rm c)}{=} \!\!\int_{0}^{\infty} \!\!f_k(r)\exp\!\Bigg(\!\!\!-2\pi\lambda_k \!\int_{r}^{\infty}\!\!\Bigg(1 -\frac{1}{\left(1+\frac{\theta}{G_k}\left(\frac{r}{t}\right)^{\alpha_k}\!\right)^{b}} \Bigg)t{\mathrm d}t \!\!\Bigg) \nonumber \\
&\times \prod_{i \in [K]^{!}} \Bigg[e^{-\lambda_i \pi(P_{ik})^{\delta_i}r^{\alpha_k \delta_i}} \nonumber \\
&\times \exp\!\left(\!-2\pi\lambda_i\int_{r^{\frac{\alpha_k}{\alpha_i}}(P_{ik})^{\frac{1}{\alpha_i}} }^{\infty} \Bigg(1-\frac{1}{\left(1+\frac{\theta P_{ik}r^{\alpha_k}}{t^{\alpha_i}}\right)^b}\Bigg)t{\rm d}t\right)\!\!\Bigg]{\rm d}r \nonumber \\
&=\int_{0}^{\infty} 2\pi\lambda_k r \exp\Bigg[-\lambda_k\pi r^2 \nonumber \\
&-2\pi\lambda_k\int_{r}^{\infty}\Bigg(1 -\frac{1}{\left(1+\frac{\theta}{G_k}\left(\frac{r}{t}\right)^{\alpha_k}\right)^{b}} \Bigg)t{\mathrm d}t  \nonumber \\
& -\sum_{i \in [K]^{!}}\lambda_i \pi(P_{ik})^{\delta_i}r^{\alpha_k \delta_i} \nonumber \\
&-2\pi\lambda_i\int_{r^{\frac{\alpha_k}{\alpha_i}}(P_{ik})^{\frac{1}{\alpha_i}} }^{\infty} \Bigg(1-\frac{1}{\left(1+\theta P_{ik}\frac{r^{\alpha_k}}{t^{\alpha_i}}\right)^b}\Bigg)t{\rm d}t \Bigg]{\rm d}r \nonumber \\
&\overset{\rm (d)}{=} \int_{0}^{\infty} 2\pi\lambda_k r \exp\bigg(-\lambda_k \pi r^2 F(b,\delta_k,\theta/G_k) \nonumber \\
&- \sum_{i \in [K]^{!}}\lambda_i \pi P_{ik}^{\delta_i}r^{\alpha_k \delta_i} F(b, \delta_i,\theta)\bigg){\rm d}r \nonumber \\
&\overset{\rm (e)}{=} \!\!\int_{0}^{\infty}  \!\!\exp\!\bigg(\!\!-s F(b,\delta_k,\theta/G_k) - \!\!\sum_{i \in [K]^{!}}\!\rho_{ik}s^{\frac{\alpha_k}{\alpha_i}} F(b, \delta_i,\theta)\bigg){\rm d}s, 
\end{align}
where $f_k(r) = 2\pi \lambda_k r e^{-\lambda_k \pi r^2}$ is the distribution of the distance of the typical user to the nearest BS that belongs to the $k$th tier, $\rho_{ik} \triangleq \frac{\lambda_i \pi P_{ik}^{\delta_i}}{(\lambda_k \pi)^{\alpha_k/\alpha_i}}$, and $F(b, \delta, \theta) \triangleq \:_2F_1(b, -\delta; 1-\delta; -\theta)$. $\mathrm{(a)}$ follows from the asymptotically exact AMAPPP approximation of $\Phi_k$ where the SIR threshold $\theta$ is shifted to $\theta/G_k$ and the point process $\Phi_k$ is replaced by a PPP denoted by $\Phi_k^{\rm PPP}$. $\mathrm{(b)}$ follows from the approximation of $\Phi_i$ by a PPP $\Phi_{i}^{\rm PPP}$. $\mathrm{(c)}$ follows from the probability generating functional (PGFL) of the PPP and averaging over the distance $\|x_0\|$ of the typical user to the nearest BS belonging to $\Phi_k^{{\rm PPP}}$. $\mathrm{(d)}$ uses
\begin{align}
\int_{1}^{\infty}\left(1-\frac{1}{(1+\theta t^{-1/\delta})^b}\right){\rm d}t \equiv \:_2F_1(b, -\delta; 1-\delta, -\theta) - 1.
\end{align}
$\mathrm{(e)}$ uses the substitution $s = \lambda_k \pi r^2 $. Finally, by summing over $[K]$, we obtain the result.


\section{Proof of Thm.~\ref{thm:G_eff}}
\label{app:G_eff}
 
Let $w_k \triangleq \frac{\lambda_k P_k^{\delta}}{\sum_{i \in [K]}\lambda_i P_i^{\delta}}$. We can then write $\hat{M}_b(\theta)$ in \eqref{eq:same_path_loss} as
\begin{align}
\hat{M}_b(\theta) = \sum_{k \in [K]} \frac{w_k}{w_kF(b, \delta, \theta/G_k) +  (1-w_k)F(b,\delta,\theta)}.
\end{align}
Noticing that for $b > 0$, $F(b, \delta, \theta/G)$ is convex in $G \in (0, \infty)$, we have
\begin{align}
\hat{M}_b(\theta) &\leq \sum_{k \in [K]} \frac{w_k}{F\left(b, \delta, \frac{\theta}{w_k G_k + (1-w_k)}\right)} \\
&\overset{({\mathrm{a}})}{\leq} \frac{1}{F\left(b, \delta, \frac{\theta}{\sum_{k\in [K]} w_k(w_kG_k + (1-w_k))}\right)} \\
& = M_b^{\rm HIP}\left(\frac{\theta}{\sum_{k\in [K]} w_k(w_kG_k + (1-w_k))}\right),
\label{eq:eff_gain_exp}
\end{align}
where $(\mathrm{a})$ is due to $\sum_{k \in [K]}w_k = 1$ and $1/F(b, \delta, \theta/G)$ being concave in $G$.

From the definition of the MISR-based gain, \eqref{eq:eff_gain_exp} can be viewed as
\begin{align}
\hat{M}_b(\theta) \leq M_b^{\rm HIP}(\theta/G_{\rm eff}),
\end{align}
where 
\begin{align}
G_{\rm eff} \triangleq \sum_{k\in [K]} w_k(w_kG_k + (1-w_k))
\end{align}
is termed the {\em effective gain} for $K$-tier HCNs.

For $b < 0$, we just need to reverse the inequality, {\em i.e.}, replace `$\leq$' by `$\geq$.'
\bibliographystyle{ieeetr}
\bibliography{paper}

\end{document}